\documentclass[a4paper,fleqn,usenatbib]{mnras}
\usepackage{multirow}
 \usepackage{lscape}
 \usepackage{array, booktabs, makecell}
\usepackage{xtab}
\usepackage{wrapfig}
\usepackage{float}
\usepackage{newtxtext,newtxmath}
\usepackage[T1]{fontenc}
\usepackage{ae,aecompl}
\usepackage{enumitem}
\usepackage{adjustbox}
\usepackage{rotating}% Including figure files
\usepackage{amsmath}	% Advanced maths commands
\usepackage{amssymb}	% Extra maths symbols

\newcommand{\astrosat}{{\it AstroSat~}}

\def\procspie{{\it Proc. SPIE}\,\,}
\def\apjs{{\it Astrophys.~J.~Suppl.}}
\def\cs{{\it Curr.~Sci.}}
\def\pasj{{\it Publ. Astron. Soc. Jpn}}
\def\asr{{\it Adv.~Space.~Res.}} 
\def\aapt{{\it Astron.~Astrophys.~Trans.}}
\def\apj{{\it Astrophys.~J.}\,}
\def\aj{{\it Astronom.~J.\,}}
\def\apjl{{\it Astrophys.~J.~Lett.}}
\def\apss{{\it Astrophys. Space Sci.}}

\def\japa{{\it J.~Astrophys.~Astron.} }

\def\mnras{{\it Mon.~Not.~R.~Astron.~Soc.} }
\def\nat{{\it Nature} }

\def\fcp{{\it Fundam.~Cosm.~Phys.} }
\def\aap{{\it Astron.~Astrophys.}\,}
\def\ac{{\it Astron.~Comput.}\,}

\def\aapss{{\it Astron.~Astrophys. Suppl.~Ser.}}
\def\apjss{{\it Astrophys.~J. ~Suppl.~Ser.}}

\def\pasp{{\it Publ. Astr. Soc. Pac.}\,}
\def\basi{{\it Bull. Astron. Soc. India.}\,}
\def\araa{{\it Annu. Rev. Astron. Astrophys.}\,}
\title[UVIT observations of NGC 2336]{A study of the star forming regions in the spiral galaxy NGC 2336 using the Ultraviolet Imaging Telescope (UVIT)}
\author[Rahna et al.]{P.~T.~Rahna,$^{1}$\thanks{E-mail: 7rehanrenzin@gmail.com}
M. Das,$^{2}$
Jayant Murthy,$^{2}$ 
S.~B.~Gudennavar$^{1}$
and S. G. Bubbly$^{1}$
\\
$^{1}$Department of Physics, Christ, Bengaluru 560029, India\\
$^{2}$Indian Institute of Astrophysics, Bengaluru 560034, India
}
\pubyear{2018}

\begin{document}
\label{firstpage}
\pagerange{\pageref{firstpage}--\pageref{lastpage}}
\maketitle

% Abstract of the paper
\begin{abstract}
We present a far-UV (FUV) and near-UV (NUV) imaging study of recent star formation in the nearby spiral galaxy NGC 2336 using the Ultraviolet Imaging Telescope (UVIT). NGC 2336 is nearly face-on in morphology and has a multi-armed, branching spiral structure which is associated with star forming regions distributed mainly along the spiral arms and the co-rotation ring around the bar. We have identified 72 star forming knots in the disk, of which only two are in the inter-arm regions and 6 in the co-rotation ring. We have tabulated their positions and estimated their luminosities, sizes, star formation rates, colors, ages and masses. The ages and masses of these star forming knots were estimated using the Starburst99 stellar evolutionary synthesis models. The star forming knots have FUV and NUV mean sizes of 485 pc and 408 pc respectively and mean stellar masses of 9.8 $\times 10^{5}$ M$_{\odot}$ that range from 5.6 $\times 10^{5}$ to 1.1 $\times 10^{6}$ M$_{\odot}$. Their star formation rates vary from 6.9 $\times 10^{-4}$ to 2.2 $\times 10^{-2}$ M$_{\odot}$/yr in NUV and from 4.5 $\times 10^{-4}$ to 1.8 $\times 10^{-2}$ M$_{\odot}$/yr in FUV. The FUV-NUV colour of the  knots is found to be bluest in the central region and becomes progressively redder as the radius increases. Our results suggest that star formation in disks with spiral structure is driven by the spiral density wave and is best traced by UV imaging as it encompasses clusters spanning a wide range of star forming ages and stellar masses.
\end{abstract}

%%%%%%%%%%%%%%%%% BODY OF PAPER %%%%%%%%%%%%%%%%%%
\section{Introduction}
Star formation is one of the most important processes governing the formation and evolution of galaxies across cosmic time \citep{Kennicutt2012}. The distribution of star-forming regions in galaxy disks and their association with large-scale structures such as spiral arms and bars can reveal how star formation is triggered in galaxies \citep{Bastian2012, Adamo2017, Chandar2017} as well as how the star clusters evolve over time \citep{Adamo2015}. The distribution of star-forming regions over spiral arms is an important way to probe the spiral structure and the spurs associated with it \citep{kim2002formation, shetty2006}. A good example is the grand design spiral M51 that has knots of star formation associated with the arms \citep{Vogel1988, Marston1993, calzetti2005, miralles2011, gutierrez2011}. Bars, circumnuclear rings, and corotation rings are also important sites of star formation that are associated with global instabilities in galaxy disks \citep{Sheth2002}.

Star formation in galaxies is measured using the star formation rate (SFR) or the mass of stars formed per unit year. \cite{searle1973} and \cite{ tinsley1968, tinsley1972} derived quantitative expressions for SFR from evolutionary synthesis models of galaxies and this led to the first prediction of the evolution of SFR with the cosmic look back time \citep{madau1996high, madau1998star, blain1999history}. With the development of more precise and direct SFR diagnostics, there are now several different star formation tracers covering a wide range of wavelengths that can be used to estimate the SFR of galaxies. The H$\alpha$ and Ultraviolet (UV) luminosity of galaxies are the most commonly used tracers of massive star formation. H$\alpha$ traces the most massive O type and early type B stars of mass greater than 17$M_{\odot}$ and measures recent star formation which is only a few million years old. The UV emission traces older stars that have ages from a few million years to 10$^{8}$ yrs \citep{lee2009, Donas1984, Schmitt2006}, as well as massive A and B-type stars (M $\geq$ 3$M_{\odot}$). We can also use indirect tracers such as the X-ray continuum \citep{ranalli2003, symeonidis2011}, nebular recombination lines (mainly the HII and OII lines) \citep{cohen1976,kennicutt1983,Kewley2004}, infrared continuum fluxes \citep{harper1973,rieke1978,telesco1980, roussel2001, Kewley2002, Kennicutt2007, Boquien2010}, the radio continuum \citep{Yun2001, Murphy2011}, and near-infrared polycyclic aromatic hydrocarbon (PAH) luminosity \citep{ roussel2001, calzetti2007}. The study of individual extragalactic H II regions in nearby galaxies is important to determine the current  SFRs, spatial distributions and elemental abundances \citep{zaritsky1994, roy1996, van1998spectroscopy, dutil1999, kennicutt2003, bresolin2005vlt, bresolin2009extragalactic}.

The UV flux is one of the best tracers of the current star formation in galaxies \citep{Kennicutt1998} provided there is only limited dust present in the system, as dust can produce significant obscuration \citep{calzetti2001dust, salim2005, burgarella2005star, overzier2010dust}. The far-UV (FUV) region is an ideal window to detect and study the massive, hot young stars like O, B type stars, blue horizontal branch stars and white dwarfs (WDs) \citep{thilker2007}. They are the principal locations for the synthesis of heavy elements \citep{kinman2007, Bianchi2011}. The near-UV (NUV) emission traces the less massive stars such as A-type stars, which are
more common. Hence, NUV emission generally has a larger spatial extent compared to FUV \citep{boissier2008, carter2011spatial}. The spatial distribution of FUV and NUV emission is, therefore, a good indicator of the different types of star formation in galaxy disks and its effect on galaxy evolution \citep{Calzetti2015}.

The UV emission in galaxies can also arise from active galactic nuclear (AGN) activity. The UV emission from the AGN is an important component of the UV luminosity from spiral galaxies \citep{makishima1994, colina1997nuclear}. Most of the AGN UV flux is emitted from the accretion disc but some may be associated with radio jets \citep{kollatschny2006}. The UV emission may also be due to nuclear star formation and jet-induced star formation \citep{shastri2007}. Hence UV imaging surveys are important to understand the connection between AGN activity and star-forming galaxies \citep{heckman2005}.

It is difficult to resolve individual stars and star-forming regions in distant galaxies. Only very nearby bright galaxies provide the opportunity to study star-forming regions in detail. In this paper, we have used the recently launched Ultraviolet Imaging Telescope (UVIT) instrument \citep{Kumar2012, tandon2017uvit} on-board the \astrosat multi-wavelength mission \citep{agrawal2006, Singh2014, agrawal2017} to do a deep study of the star-forming regions in the nearby spiral galaxy NGC~2336. Our aim is to characterize the spatially resolved star-forming knots in the disk of this galaxy, determine their SFRs and study their properties. The UVIT has a high angular resolution of 1$^{\prime\prime}$ \citep{Rahna2017}, as compared to the previous dedicated UV survey mission GALEX, which has an angular resolution of 4.3$^{\prime\prime}$-5.3$^{\prime\prime}$ \cite{Morrissey2007}. Thus UVIT is ideal for studying star-forming regions in very nearby face-on galaxies, where it can resolve large star clusters associated with the spiral arms. High angular resolution and multiple filters in the UV waveband are the unique features of UVIT compared to GALEX.

The paper is organized as follows: section 2 gives a description of the sample galaxy, section 3 describes the observations and the data reduction. The results and discussion are presented in section 4. A summary of our conclusions is given in section 5.
%%%%%%%%%%%%%%%%%%%%%%%%%%%%%%%%%%%%%%%%%%%%%%%%%%%%%%%%%%
\section {The galaxy-NGC 2336}
\begin{table*}
\centering
\caption{Properties of NGC 2336}
\label{ngc2336_para}
\begin{tabular}{lll}
\hline
\textbf{Parameter} & \textbf{Value} & \textbf{Reference} \\ \hline
RA &  07:27:04.05 & NED data base\\
DEC & +80:10:41.1 & NED data base \\
Type & SAB(r)bc & NED data base \\
Total apparent B magnitude ($m_{B}$) & 11.19 & \cite{gusev2016spectral} \\
Absolute B magnitude ($M_{B}$) & -22.14 & \cite{gusev2003structure} \\
Radial velocity with respect to the Local Group ($V_{LG}$) & 2417.7 km/s & NED data base \\
Distance (Galactocentric GSR)(R) & 32.2 Mpc & NED data base\\
Scale (Galactocentric GSR)& 156 pc/arcsec & NED data base \\
Apparent diameter ($D_{25}$) (arcmin) & 6.31& \cite{gusev2016spectral} \\
Apparent corrected diameter ($d_{25}$) & 64.9 kpc & \cite{gusev2016spectral} \\
Inclination (i) & 50.95$^{\circ}$ & NED data base \\
Position angle (PA) & 155$^{\circ}$ & NED data base\\ 
Redshift(z)&0.00735 & NED data base \\ 
Major Diameter (arcmin) & 7.1 &  NED data base \\
Minor Diameter (arcmin) & 3.9 & NED data base\\\hline
\end{tabular}
\end{table*}
\begin{figure*}
\centering
\includegraphics[scale=0.4]{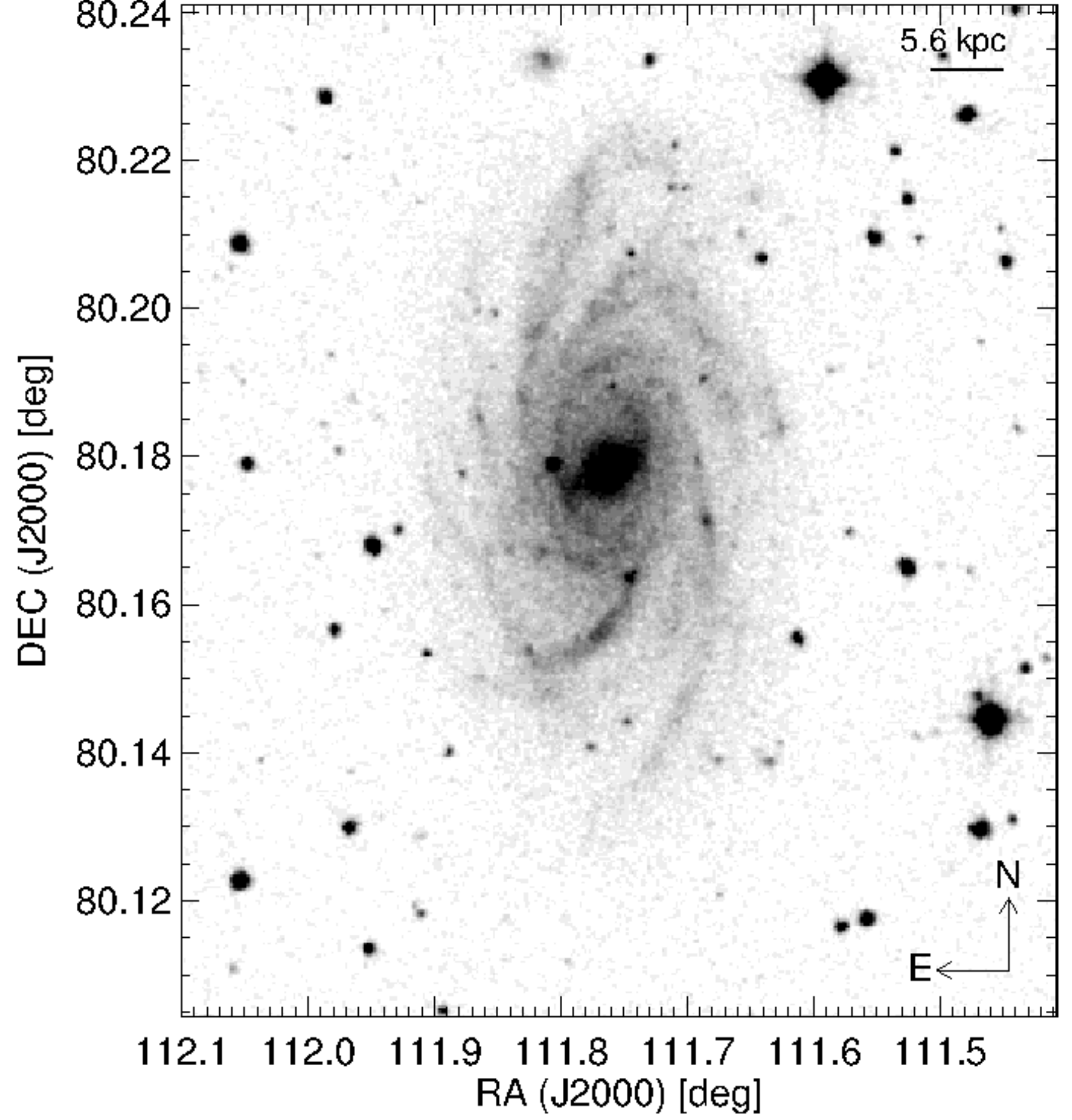}
\includegraphics[scale=0.4]{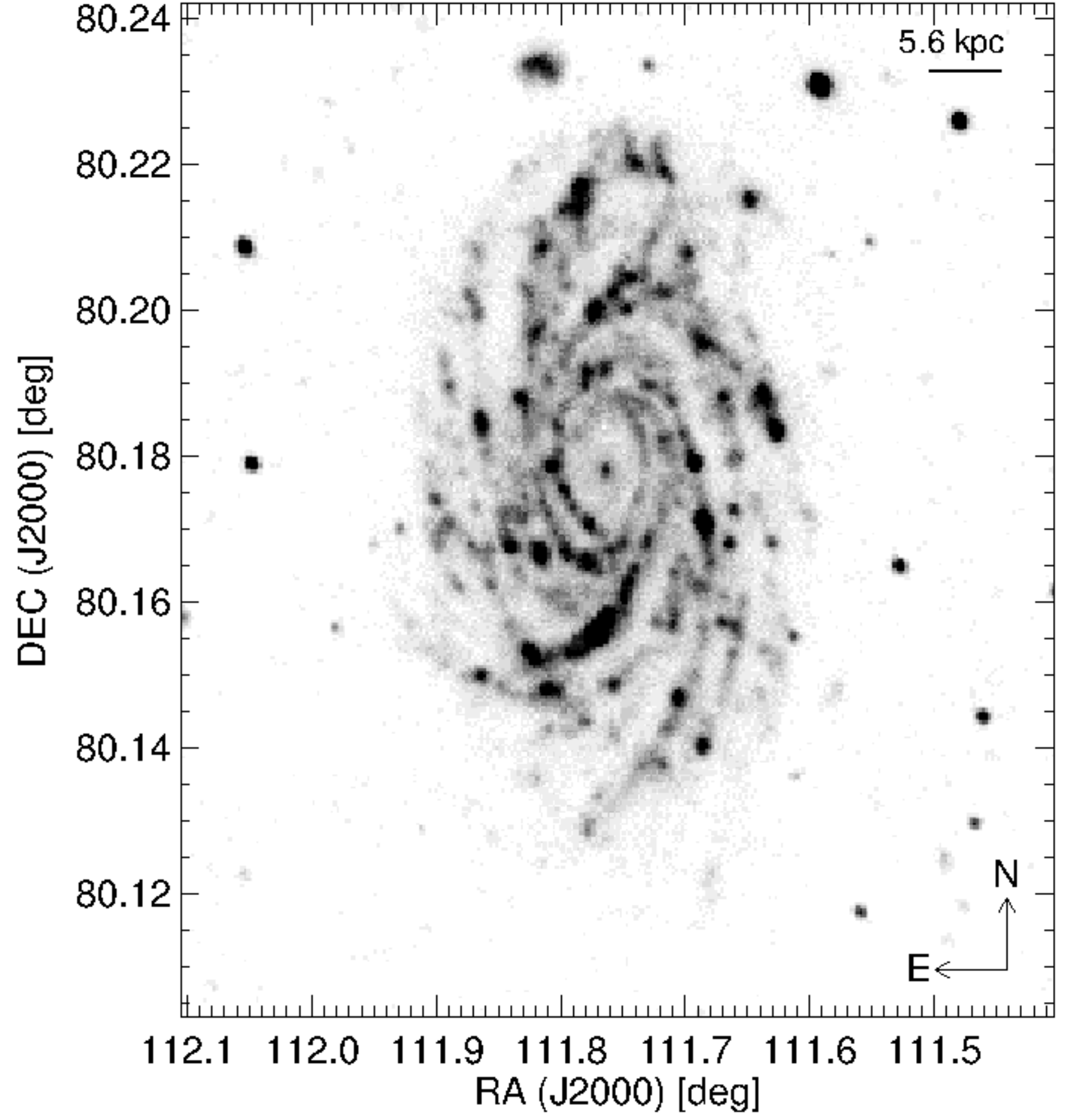}
\caption{The optical (DSS-R: left) and UV (GALEX-NUV: right) image of NGC 2336.}
\label{fig:ngc2336_vis_uv}
\end{figure*}
NGC 2336 (UGC 3809) is a barred ringed spiral galaxy, located in the constellation Camelopardalis (Table \ref{ngc2336_para}). The optical and UV images of NGC 2336 clearly show that it is close to face-on and has a multi-armed spiral structure (Figure~\ref{fig:ngc2336_vis_uv}). It has the Hubble classification of SAB(r)bc type. It is one of the few galaxies that is thought to strongly resemble our Milky way galaxy \citep{efremov2011}. At a distance of 32.2 Mpc, the galaxy scale length is 156pc/arcsecond, which is small enough to be able to resolve the star-forming regions in the disk as well as trace morphological features such as bars, rings, and spiral arms. NGC2336 is part of a non-interacting pair with the galaxy IC 0467, which is at a distance of  20$^{\prime}$ or $\sim$187~kpc\citep{van1983neutral} and does not appear to influence the structure and kinematics of the galaxy \citep{schneider1992velocity, keel1993kinematic}.

NGC~2336 is a bright galaxy and has prominent star forming regions associated with its spiral arms. The multi-arm and branched spiral structure is the most striking morphological feature of this galaxy. Young stars, gas, and dust make up the star forming regions along the arms \citep{gusev2003structure}. The disk has six well developed spiral arms arising from the ring around the small bar \citep{boroson1981,biviano1991}. There are four spiral arms in the northern and western regions, a bright arm in the southern region and a faint arm in the eastern part of the galaxy \citep{gusev2003structure}.

The galaxy has a very small, star-like nucleus that is $\sim$0.8 kpc in size. Since no strong nuclear emission has been detected in either the optical \citep{Heckman1980,tifft1982}, IR \citep{quillen2001} or radio \citep{Heckman1980, hummel1985, niklas1995radio} wavelengths, the nucleus of this galaxy is classified as non-active. There is a prominent bulge of radius 12$^{\prime\prime}$ or 1.8 kpc, that has an ellipticity of 0.2 and position angle (PA) of 172$^{\circ}$ \citep{gusev2003structure}.

The contribution from the bulge to the galaxy luminosity in B and R bands is 5-6\% \citep{boroson1981, grosbol1985} and 21\% in the J band \citep{Wilke1999}. There is a short, weak bar with an apparent length scale of 24$^{\prime\prime}$ and deprojected size of 37.5$^{\prime\prime}$ or 5.9~kpc. The surface brightness of the bar is 21.4 mag arsec$^{-2}$ in B band and 20.6 mag arsec$^{-2}$ in V band  \citep{gusev2003structure}. The bar has a PA of  117$^{\circ}$  in the V, R and I bands. It has an ellipticity of 0.33 in the U, B, V and R bands. The bar is surrounded by a co-rotation ring of radius 34.1$^{\prime\prime}$ or 5.3 kpc and a mean surface brightness of 21.8 mag arsec$^{2}$ in B band \citep{gusev2003structure}. The bar contributes about 14\% of the luminosity of the galaxy in J band \citep{Wilke1999}.

The total mass of NGC 2336 within a 4\arcmin.1 radius is $4.1 \times 10^{11} M_{\odot}$  \citep{van1983neutral}. However, \cite{wilke2001mass} constructed a kinematical and dynamical 2D model of NGC 2336 by fitting a disk, a bulge, and a bar to the observed surface brightness distribution and derived its physical parameters; they estimated the total mass of NGC 2336 is $1.2 \times 10^{11} M_{\odot}$. They found the bulge mass is $1.2 \times 10^{10} M_{\odot}$, disc mass is  $9.63 \times 10^{10} M_{\odot}$ and the bar mass is $1.13 \times 10^{10} M_{\odot}$. NGC~2336 is a  gas-rich galaxy and has an HI mass of $4.57 \times 10^{10} M_{\odot}$ \citep{wunderlich1991radio, young1996, martin1998catalogue}, molecular gas ($H_{2}$) mass of $1.82 \times 10^{10} M_{\odot}$  \citet{young1996}. The dust mass is $2.14 \times 10^{6} M_{\odot}$ \citet{young1996}. The HI in this galaxy has been mapped in many surveys \citep{shostak1978, dickel1978, rots1980}. The HI map shows a lack of neutral hydrogen in the central parts of this galaxy \citep{van1983neutral}.

There are HII regions giving rise to H$\alpha$ emission all along the spiral arms \citep{Garrido2004}. It has high H$\alpha$ luminosity of $ 7.41 \times 10^{8} L_{\odot}$ \citep{young1996} and the FIR luminosity is $3.09 \times 10^{10} L_{\odot}$. NGC~2336 has been observed in the optical \citep{boroson1981, blackman1984, grosbol1985, heraudeau1996, baggett1998, gusev2003structure, gusev2012oxygen, gusev2016spectral, gusev2016study}, infrared \citep{ aaronson1982, wunderlich1991radio, keel1993kinematic, van1993iras, Wilke1999, Font2017} and radio \citep{Heckman1980, aaronson1982, van1983neutral, young1995, niklas1995radio}. For this study, we use the prominent UV emission from the disk to study star forming regions along the spiral arms and the corotation ring in detail. 
%%%%%%%%%%%%%%%%%%%%%%%%%%%%%%%%%%%%%%%%%%%%%%%%%%%%%%%
\section{UVIT Observations and Data}
\begin{table*}
\centering
\caption{Filters used in observations}
\label{obs2336}
\resizebox{\textwidth}{!}{
\begin{tabular}{lllll|llll}
\hline
\multirow{2}{*}{} & \multicolumn{4}{c|}{\textbf{FUV}} & \multicolumn{4}{c}{\textbf{NUV}} \\ 
 & \multicolumn{1}{l}{\textbf{CaF2\_1 (F1)}} & \multicolumn{1}{l}{\textbf{BaF2 (F2)}} & \multicolumn{1}{l}{\textbf{Sapphire (F3)}} & \multicolumn{1}{l|}{\textbf{Silica (F5)}} & \multicolumn{1}{l}{\textbf{Silica (F1)}} & \multicolumn{1}{l}{\textbf{NUVB15 (F2)}} & \multicolumn{1}{l}{\textbf{NUVB13 (F3)}} & \multicolumn{1}{l}{\textbf{NUVB4 (F5)}} \\ \hline
\textbf{Central $\lambda_0$(\AA)} & 1509.4 & 1549.6 & 1607 & 1703 & 2418 & 2185 & 2436 & 2628 \\
\textbf{Pass band (\AA)} & 1250-1790 & 1330-1830 & 1450-1810 & 1600-1790 & 1940-3040 & 1900-2400 & 2200-2650 & 2200-2650 \\
\textbf{Effective bandwidth $\Delta\lambda$ (\AA)} & 441 & 378 & 274 & 131.3 & 769 & 271 & 281.7 & 282.3 \\
\textbf{Total exposure time (s)} &1863.771  &4170.533  & 4107.885 &4780.423  &2941.2987 &3344.7563  &4803.1375  & 6424.7176  \\
\textbf{Date of observation} &17 Dec.2015 &18 Dec.2015 & 18 Dec.2015 & 18,19 Dec. 2015 & 17 Dec.2015 &18 Dec. 2015 &18 Dec. 2015  & 18,19 Dec. 2015 \\
 \hline
\end{tabular}
}
\end{table*}
\begin{figure*}
%\centering
\includegraphics[scale=0.4]
{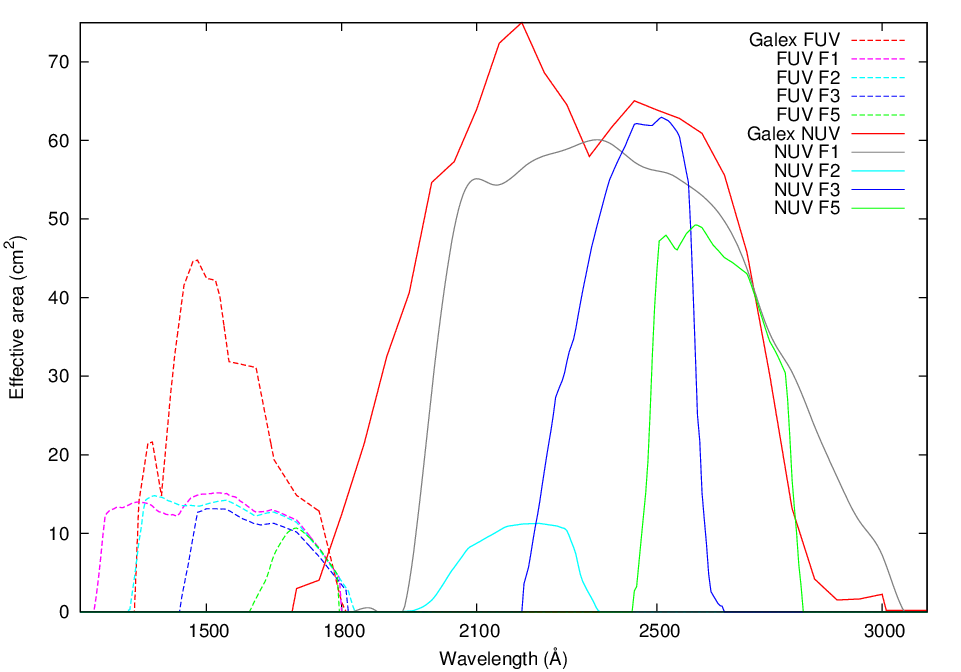}
\caption{Effective area of the filters of FUV and NUV channels used in this study with GALEX filters.}
\label{fig:filters}
\end{figure*}
The \astrosat mission was launched on 28 September, 2015 into a low Earth orbit (650 km, $6^{\circ}$ inclination) using the Polar Satellite Launch Vehicle (PSLV). There are 5 payloads on the satellite with spectral coverage from the hard X-ray (0.5 -- 100 keV) to the optical \citep{Singh2014}. We will focus on data from the Ultraviolet Imaging Telescope (UVIT) \citep{Kumar2012}. The UVIT consists of two 35-cm Ritchey-Chr\`etien telescopes with one channel dedicated to the far ultraviolet (FUV: 1250 -- 1830 \AA) and the other split between the near ultraviolet (NUV: 1900 -- 3040 \AA) and the visible (VIS: 3040 -- 5500 \AA) through the use of a dichroic beam splitter. The three detectors, one for each channel, are identical except for different photo-cathodes feeding an intensified micro channel plate (MCP) detector with a $512 \times 512$ pixel CMOS readout at 29 frames per second. The individual photon hits are recorded in each frame, centroided on-board to $\frac{1}{8}$ of a pixel and stored and transmitted to the ground. The raw spacecraft data are processed by the Indian Space Science Data Centre (ISSDC) and the Level 1 data, for each instrument, are provided to the users.

The observations of NGC 2336 were done over 23 orbits early in the performance and verification (PV) phase of UVIT. Each orbit consisted of simultaneous observations in the three channels (FUV, NUV, and VIS) through different filters for each channel with total effective exposure times given in Table \ref{obs2336}. Note that this is less than the actual time on target because some part of the exposure was lost to cosmic rays, which resulted in a huge burst of Cerenkov radiation ruining the entire frame. As this observation was taken early in the mission, there were also issues with the spacecraft guidance and with the telemetry and software which resulted in the loss of more frames.

The FUV observations were made with four crystalline filters and the NUV with three narrow-band interference filters and one broad-band crystalline filter with effective areas shown in Figure \ref{fig:filters}. We have shown the effective areas for the FUV and NUV detectors on the {\it Galaxy Evolution Explorer} ({\it GALEX}) for comparison. The VIS channel is intended primarily for spacecraft tracking and is, generally, not recommended for scientific analysis.

We have used the JUDE pipeline \citep{Murthy_ascl,murthy2017} to reduce the Level 1 data from the ISSDC into Level 2 photon lists and images for scientific analysis. The astrometric correction for the images is done using Astrometry.net \citep{lang2010} following which we placed all the data on the same coordinate reference frame and co-added those data with the same filter (Table \ref{obs2336}). The photometric calibration is taken from \cite{Rahna2017}.

%%%%%%%%%%%%%%%%%%%%%%%%%%%%%%%%%%%%%%%%%%%%%%%%%%%%%%%%%
\section{Results and discussions}
\subsection{Overview}
\begin{figure*}
\centering
\includegraphics[scale=0.4]{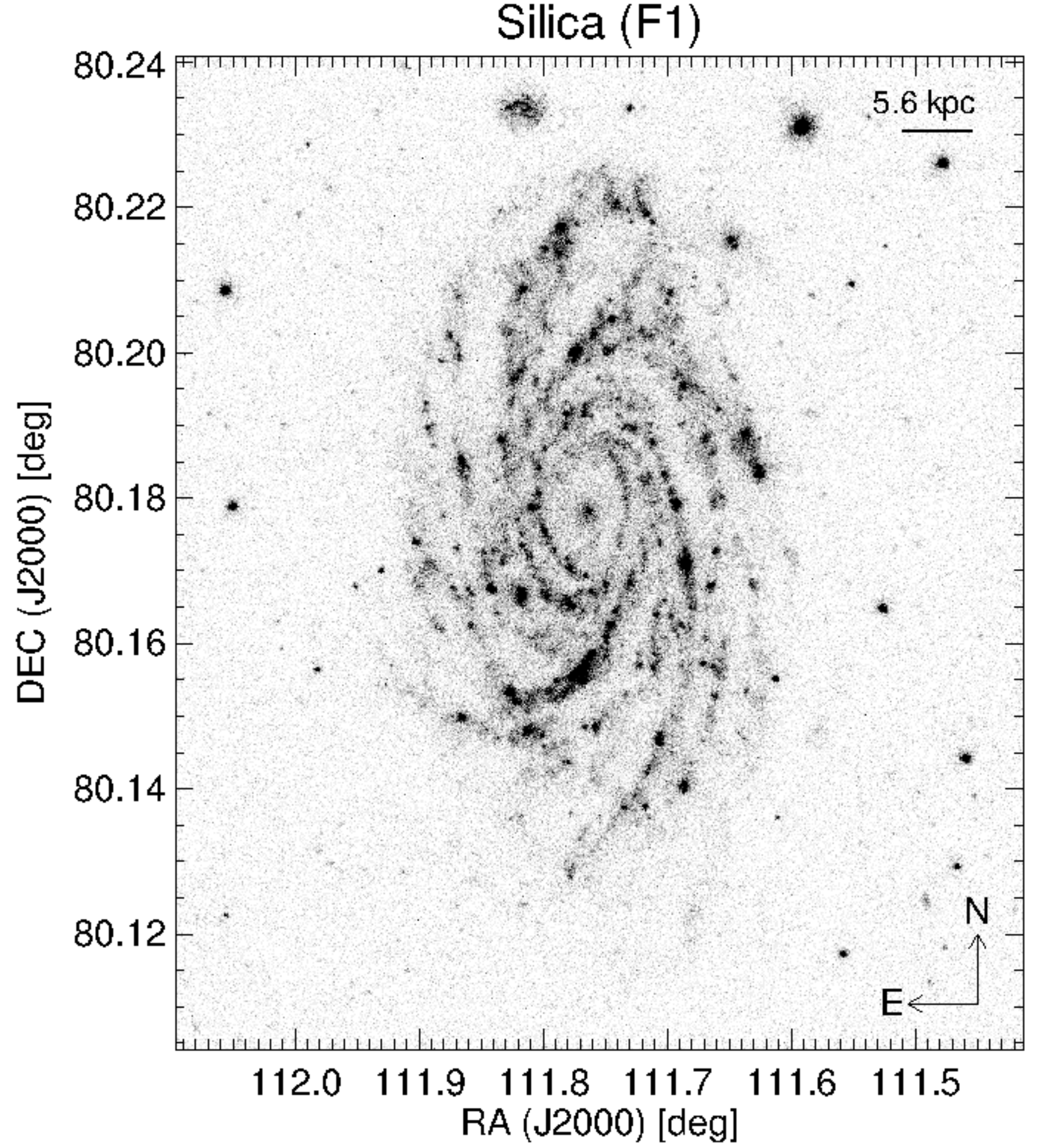}
\includegraphics[scale=0.4]{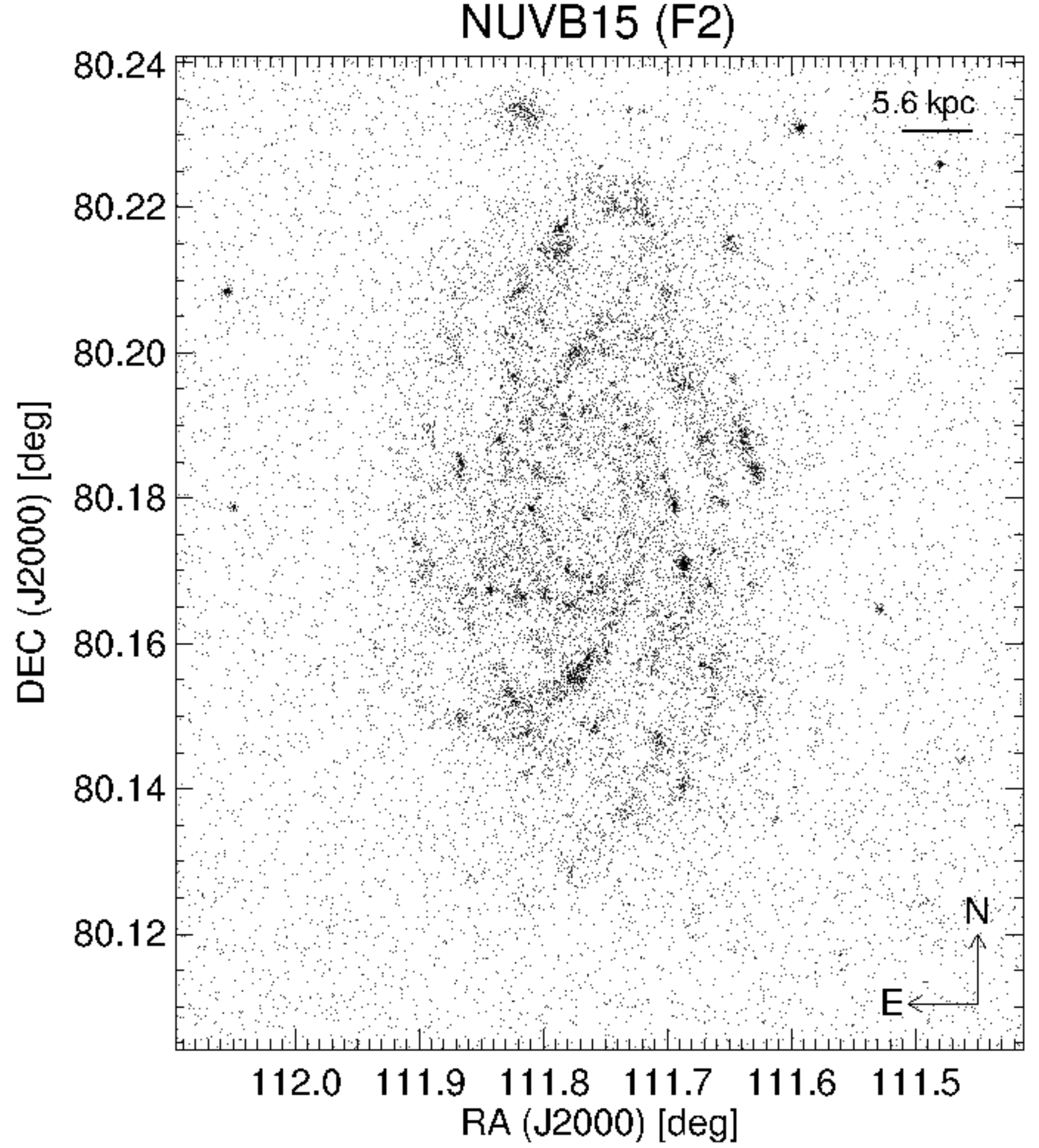}
\includegraphics[scale=0.4]{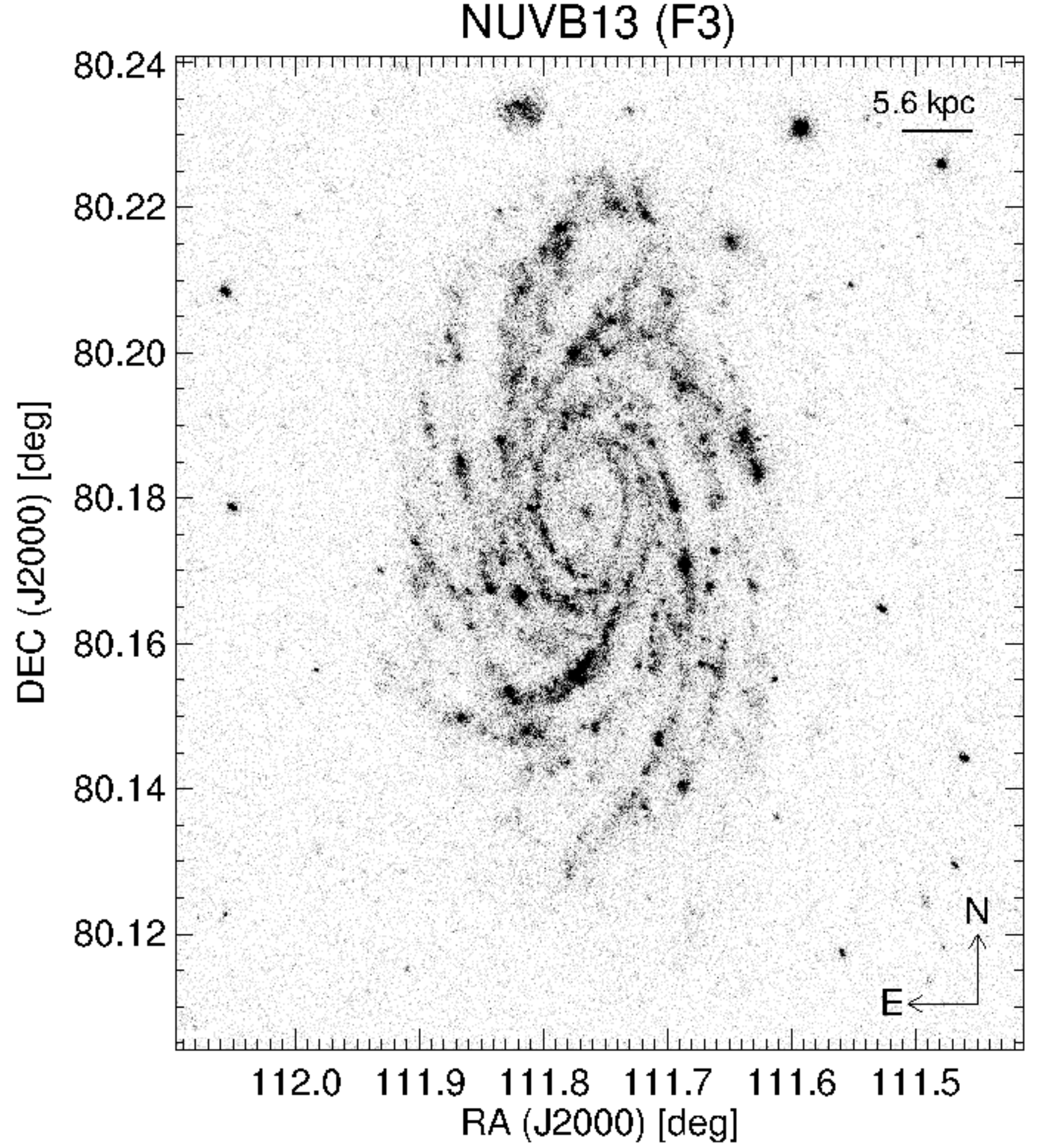}
\includegraphics[scale=0.4]{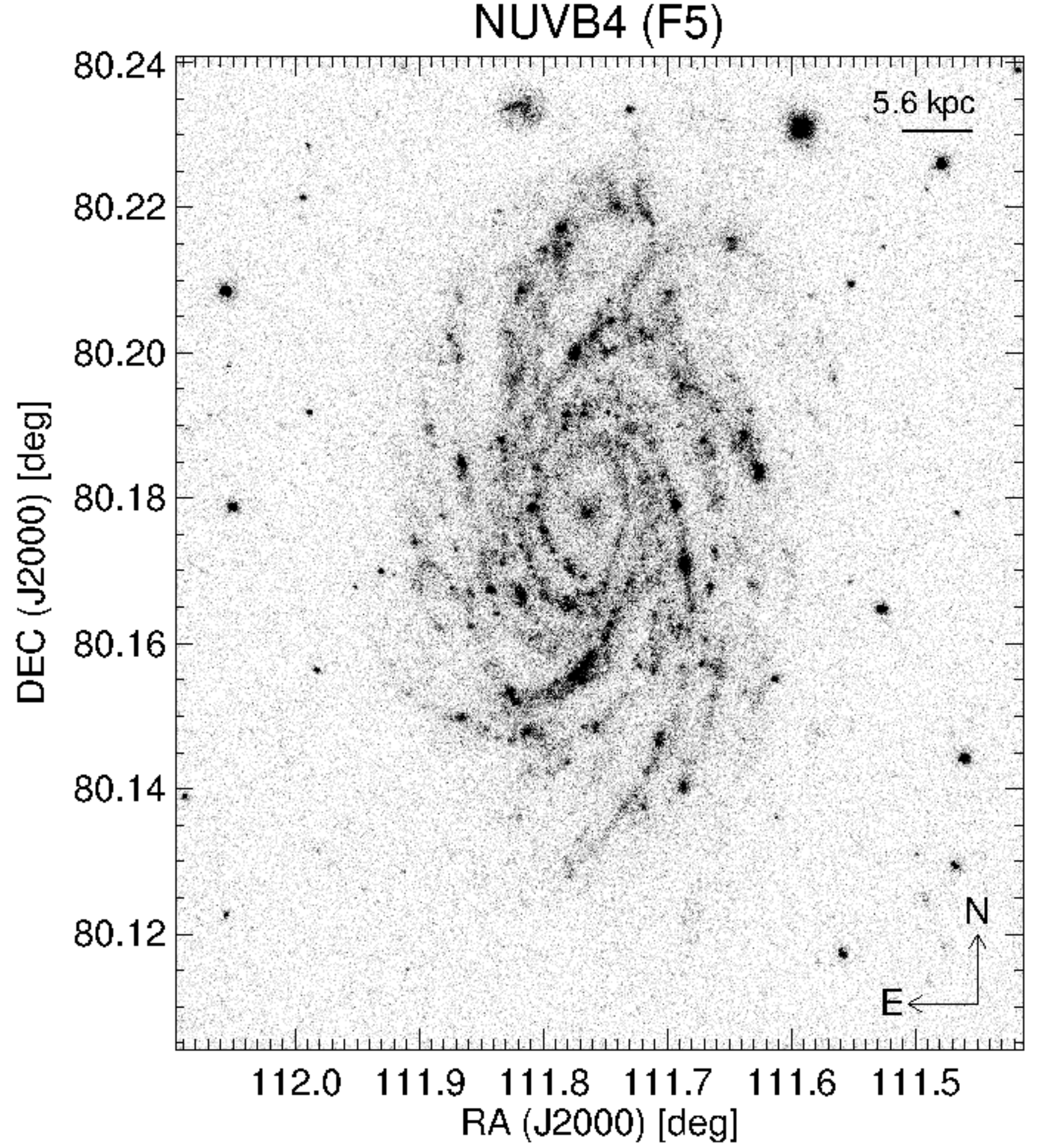}
\caption{ NUV images of NGC 2336 in different filters.}
\label{fig:Nuv}
\end{figure*}

\begin{figure*}
\centering
\includegraphics[scale=0.4]{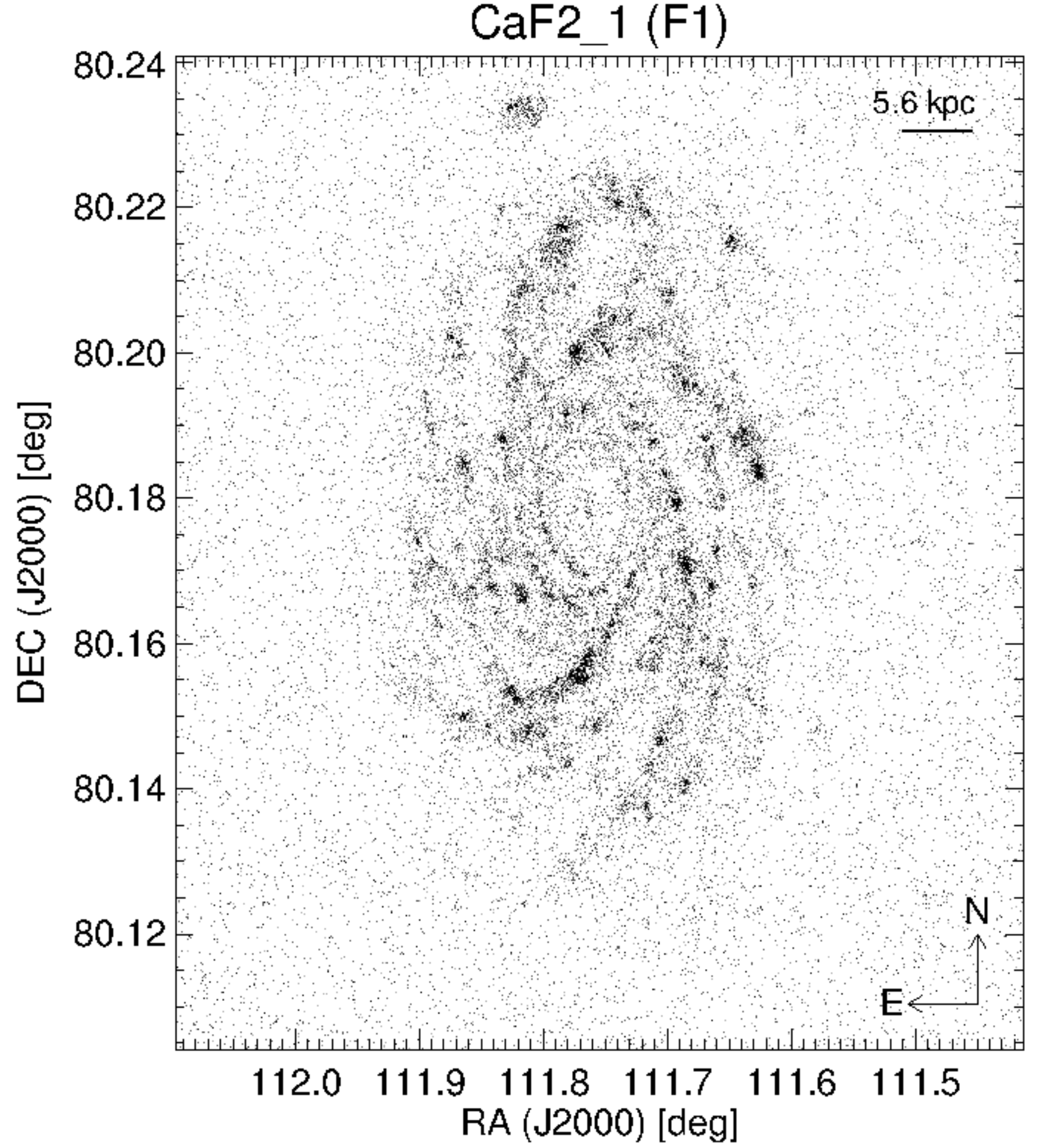}
\includegraphics[scale=0.4]{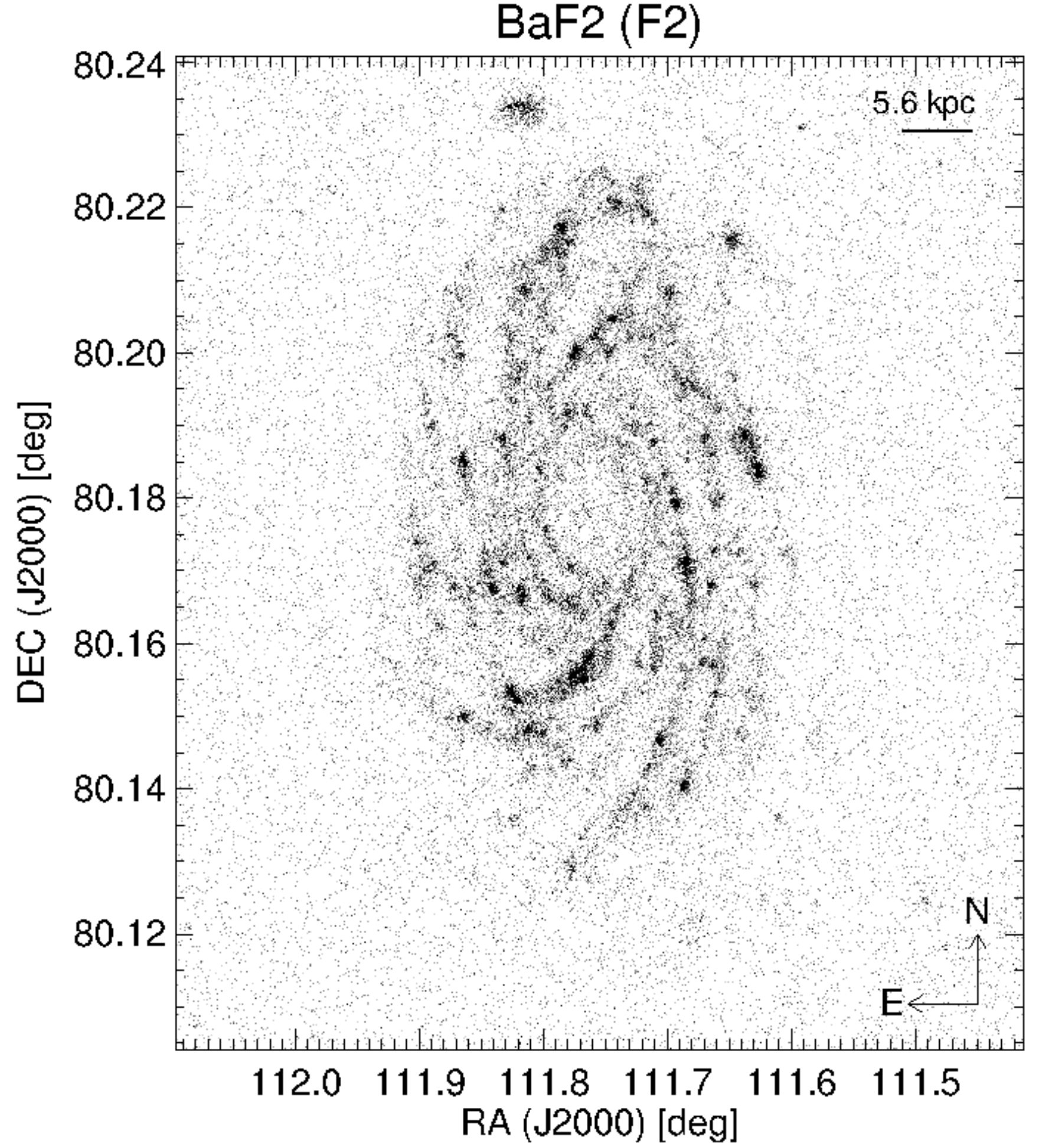}
\includegraphics[scale=0.4]{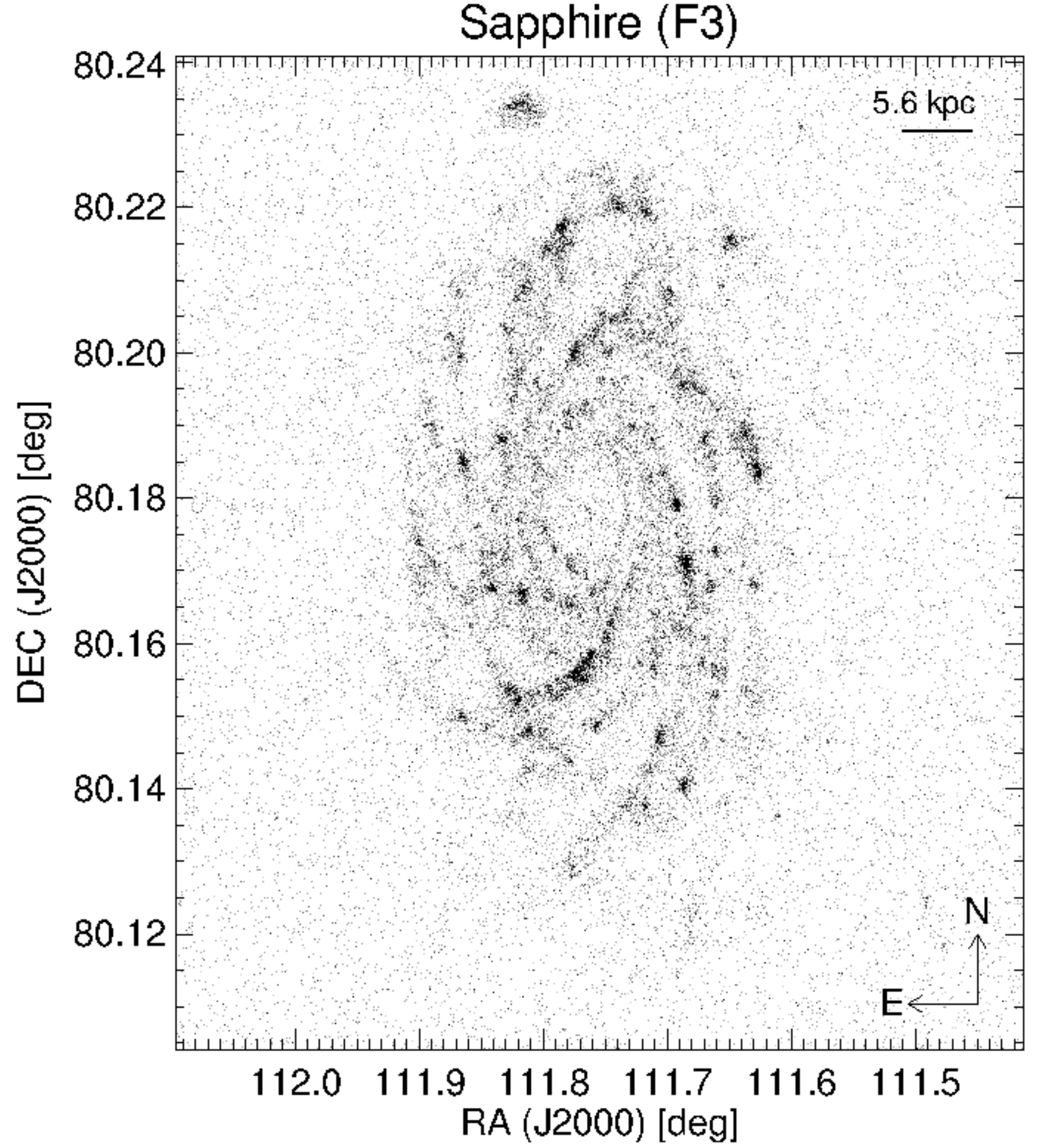}
\includegraphics[scale=0.4]{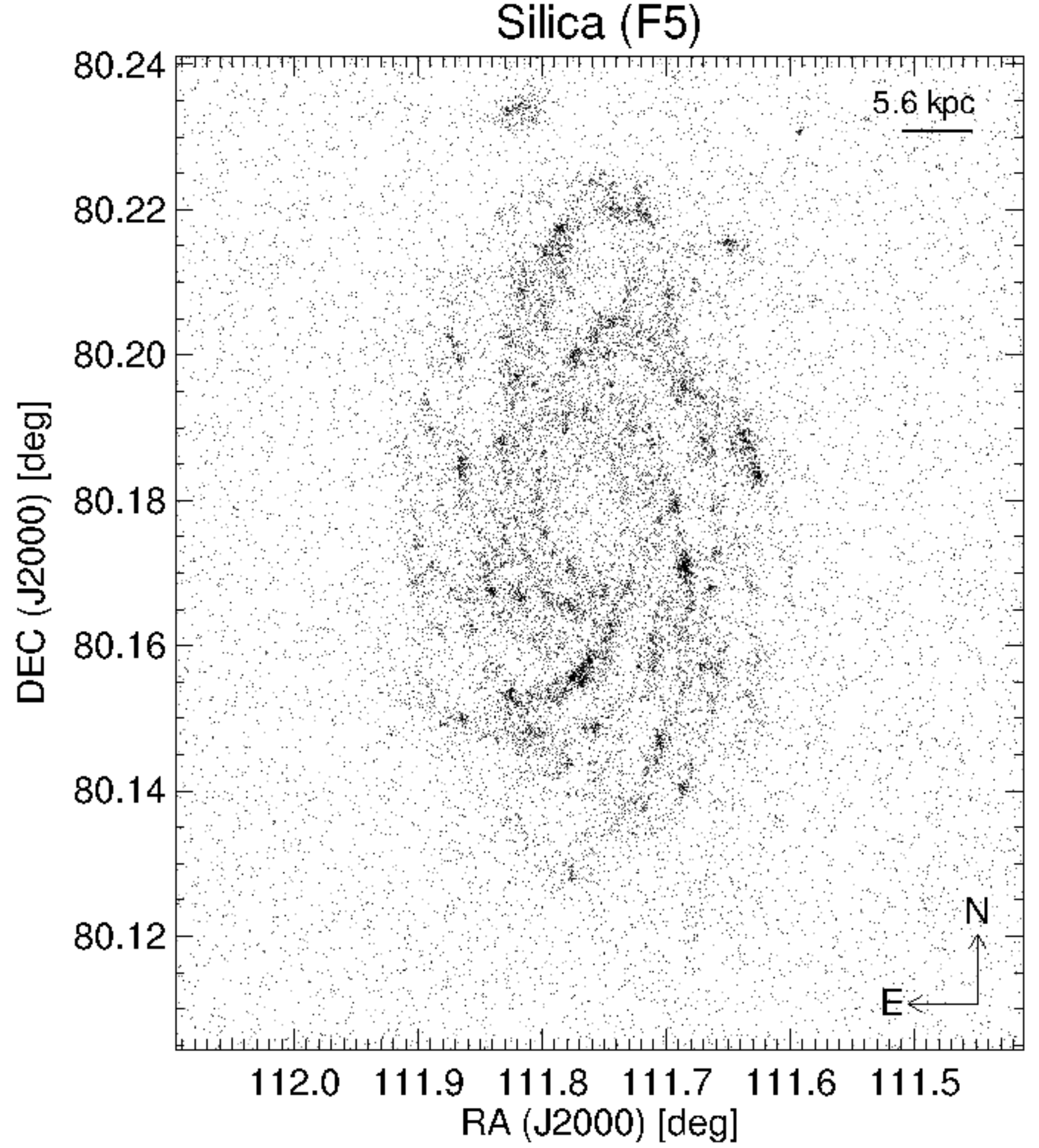}
\caption{FUV images of NGC 2336 in different filters.}
\label{fig:fuv}
\end{figure*}
\begin{figure*}
\centering
\includegraphics[scale=0.4]{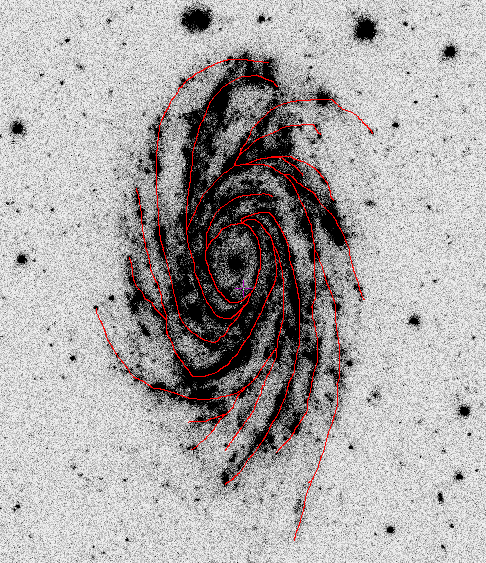}
\includegraphics[scale=0.4]{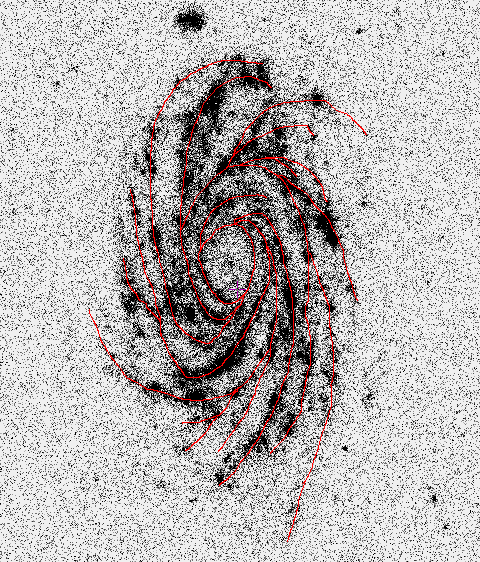}
\caption{The NUV (Left) and FUV (Right) images of UVIT with the outline of the spiral arms sketched (for visual purposes).}
\label{fig:uv_outline}
\end{figure*}
 \begin{table*}
\centering
\caption{Photometry of NGC 2336 in different filters}
\label{Tab:totalflux}
\resizebox{\textwidth}{!}{
\begin{tabular}{llllllll}
\toprule
\textbf{Filter} & \textbf{CPS} & \textbf{CPS err} & \textbf{Flux$^{a}$} & \textbf{Flux err$^{a}$} & \textbf{\thead{Area \\{[}arcsec$^{2}${]}}} & \textbf{\thead{Surface\\ brightness$^{b}$ }} & \textbf{\thead{Surface\\brightness err$^{b}$}} \\ 
\midrule
NUVF1 & 407.866 & 0.372 & 9.89E-14 & 9.03E-17 & 80374.541 & 5.05E-23 & 4.61E-26 \\
NUVF2 & 16.359 & 0.070 & 1.19E-13 & 5.08E-16 & 80374.541 & 6.08E-23 & 2.60E-25 \\
NUVF3 & 109.213 & 0.151 & 1.40E-13 & 1.94E-16 & 80374.541 & 7.18E-23 & 9.92E-26 \\
NUVF5 & 110.800 & 0.131 & 9.61E-14 & 1.14E-16 & 80374.541 & 4.91E-23 & 5.82E-26 \\
FUVF1 & 47.641 & 0.160 & 2.30E-13 & 7.73E-16 & 80374.541 & 1.18E-22 & 3.95E-25 \\
FUVF2 & 38.848 & 0.097 & 2.04E-13 & 5.07E-16 & 80374.541 & 1.04E-22 & 2.59E-25 \\
FUVF3 & 27.900 & 0.082 & 1.90E-13 & 5.60E-16 & 80374.541 & 9.70E-23 & 2.86E-25 \\
FUVF5 & 15.063 & 0.056 & 2.69E-13 & 1.00E-15 & 80374.541 & 1.37E-22 & 5.12E-25 \\
\bottomrule
\multicolumn{8}{l}{$^{a}$ erg/s/cm$^{2}$/\AA} \\
\multicolumn{8}{l}{$^{b}$ erg/s/cm$^{2}$/\AA /pc$^{2}$}
\end{tabular}
}
\end{table*}
We have observations of NGC 2336 in 8 filters with UVIT and these are shown in Figure~\ref{fig:Nuv} and Figure~\ref{fig:fuv}. The spiral structure of this galaxy is clearly discernible in the UV due to the presence of hot young stars. They are clustered in the knots of star formation along the spiral arms. The galaxy appears to be near face-on \citep{Wilke1999} making it easy to trace the distribution of star formation in the galaxy.

The galaxy NGC 2336 has a well defined multi-armed spiral structure  (Figure ~\ref{fig:uv_outline}). These multi-arms are formed due to the global instabilities in the disc of the galaxy \citep{goldreich1965, shu2016} which drives the formation of star clusters by gas compression and shocks. Although the position of the spiral arms in both the FUV and NUV images are the same, the arms are thicker in the NUV images in the outer disk regions. The branching of the spiral structure is due to ultra harmonic resonances (1:4) in the spiral density wave \citep{shu1973}. Surprisingly, the spiral arms do not originate from the ends of the bar, instead, they appear to originate from the co-rotation ring around the bar. We find that there are 3 major arms originating from the central nuclear ring. These major arms branch to the north and south, with more branching in the southern half. There are 9 minor branches originating from the southern major arm and 7 from one of the northern arms. The other northern major arm is small and has no branches. We calculated the total integrated FUV and NUV flux of this galaxy in each of the 8 filters by taking an elliptical region around the galaxy and determining the total flux and surface brightness (Table~\ref{Tab:totalflux}). 

The fluxes are measured by employing UVIT counts to flux conversions from \cite{Rahna2017} and errors were assessed by adding Poisson photon statistics. We corrected the measured flux for extinction by assuming a local Milky Way-like extinction curve using \cite{Calzetti2000}. The value of $A_{UV}$/E(B$-$V) is taken from \cite{Bianchi2011} which they calculated from the typical extinction curve for MW from \cite{Cardelli1989}. We applied this correction for all measured fluxes in this paper. 
%%%%%%%%%%%%%%%%%%%%%%%%%%%%%%%%%%%%%%%%%%%%%%%%%%%%%%%%%%%%%%%%%%
\subsection{Comparison with other wave-band images}
\begin{figure*}
\centering
\includegraphics[scale=0.5]{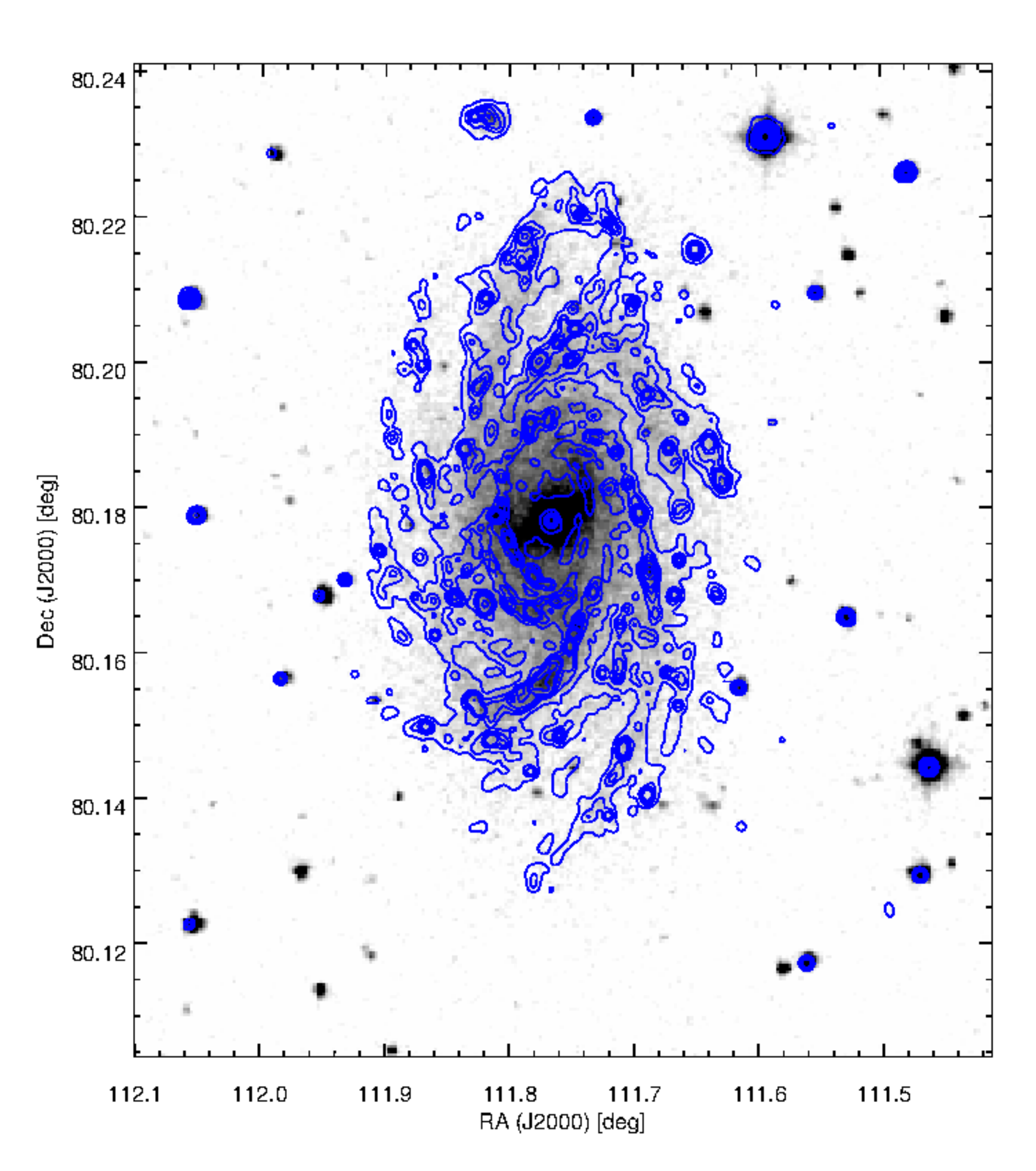}
\includegraphics[scale=0.5]{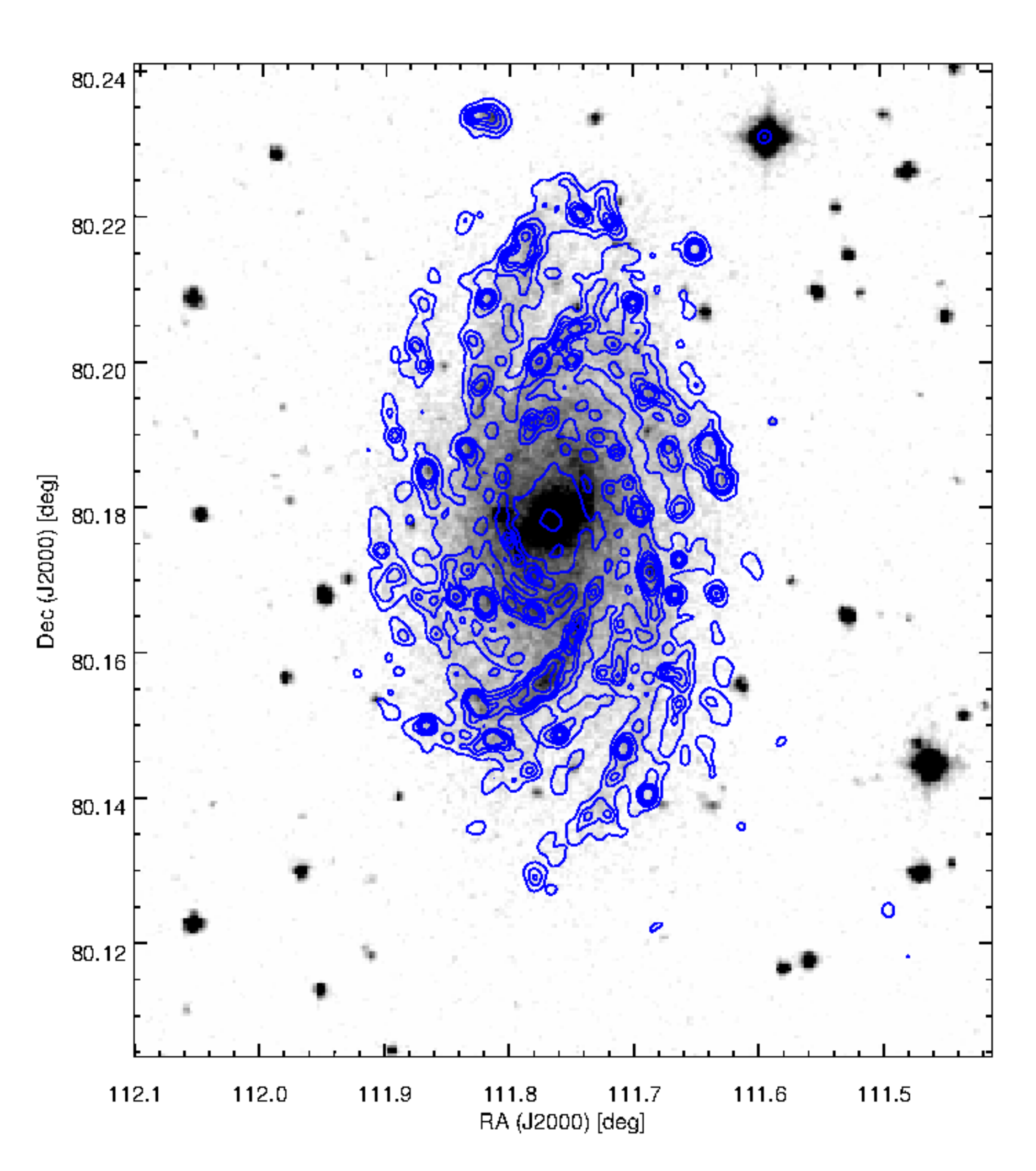}
\caption{ Overlay of UV images on DSS2R band image. Contour map of NUV in Silica filter at the top and FUV image in BaF2 filter at the bottom. The contour levels of NUV emission are taken as 5\%, 10\%, 20\%, 30\%, 40\%, 50\%, 60\%, 70\%, 80\%, 90\% and 100\% of the peak value of the count rate and contour levels of FUV emission are taken as  1\%, 2\%, 3\%, 4\%, 5\%, 10\%, 20\%, 40\%, 60\%, 80\% and 100\% of the peak value of the count rate. }
\label{fig:uv_R}
\end{figure*}
\begin{figure*}
\centering
\includegraphics[scale=0.4]{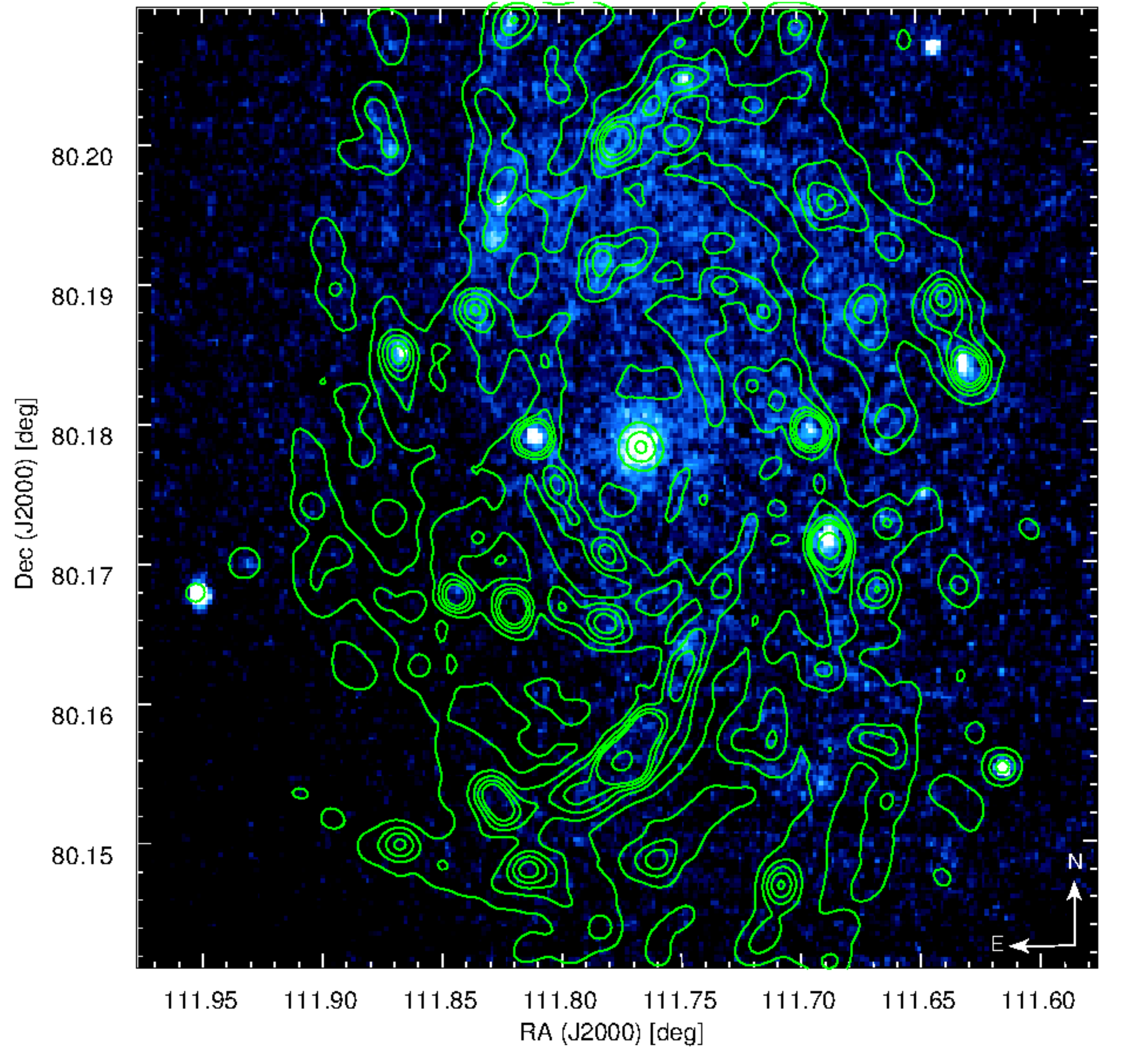}
\includegraphics[scale=0.4]{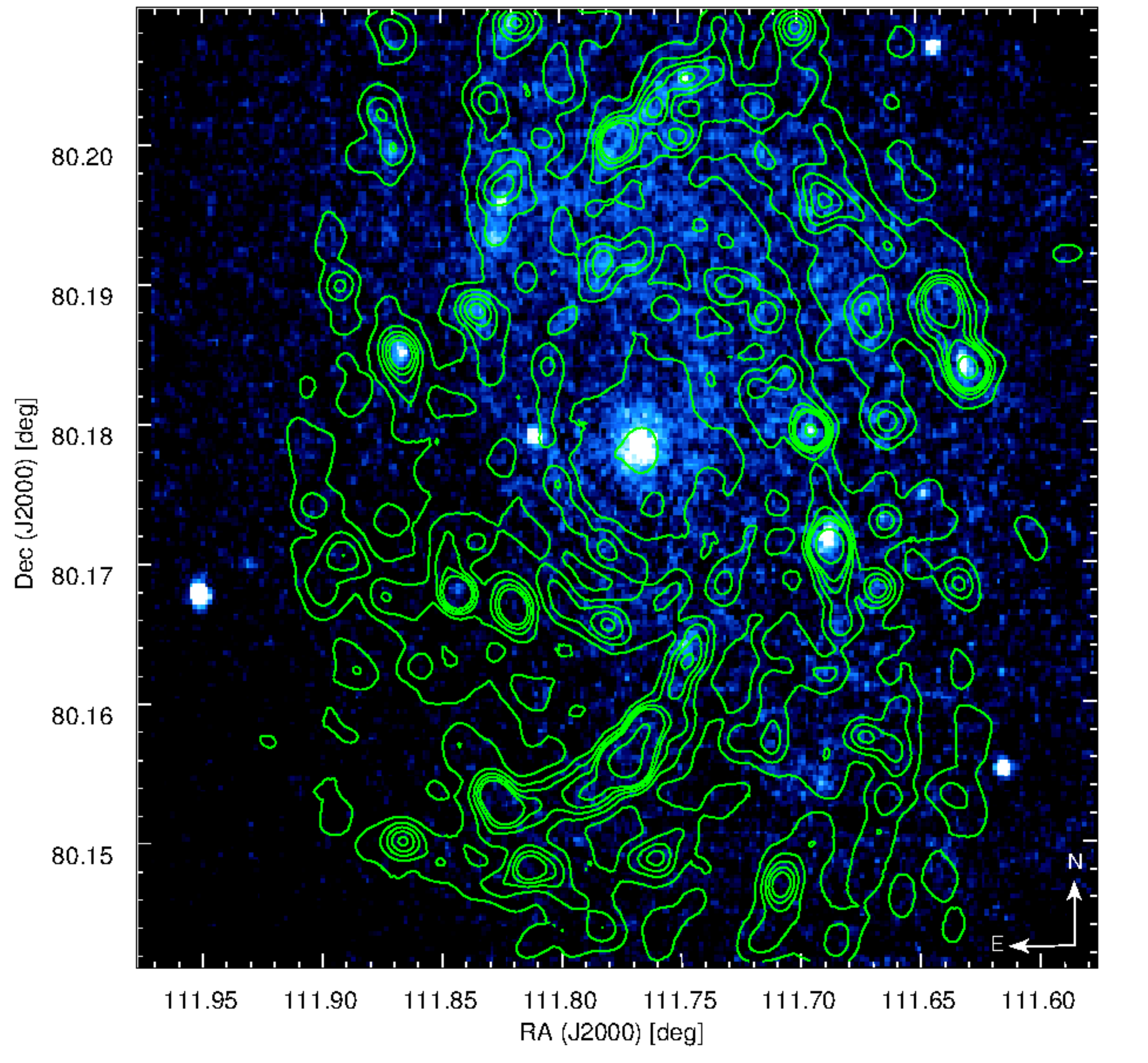}

\caption{ Overlay of UV images on H$\alpha$ image. Contour map of NUV in Silica filter at the top and FUV image in BaF2 filter at the bottom. The contour levels are as in Figure~\ref{fig:uv_R}.}
\label{fig:uv_Halpha}
\end{figure*}
\begin{figure*}
\centering
\includegraphics[scale=0.5]{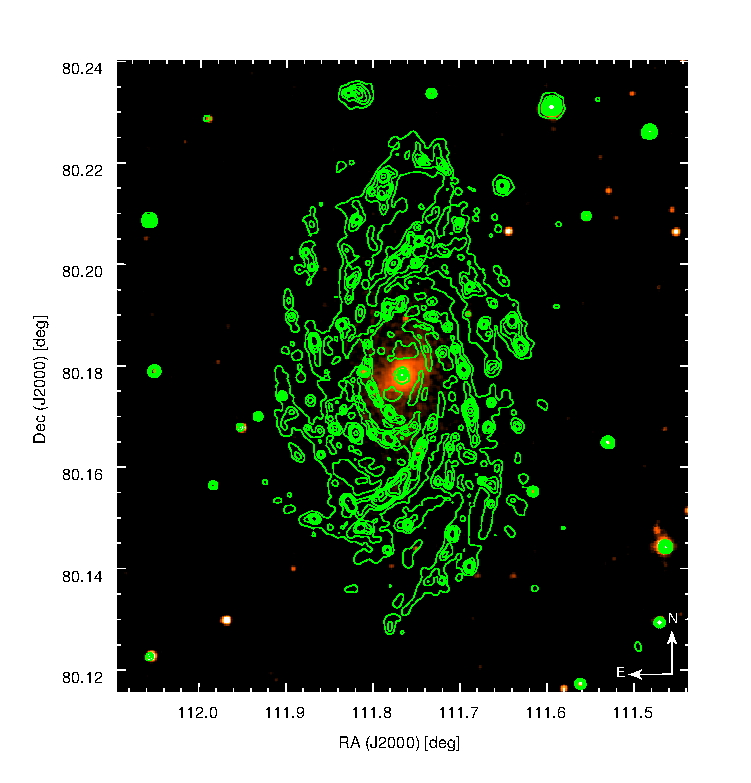}
\includegraphics[scale=0.5]{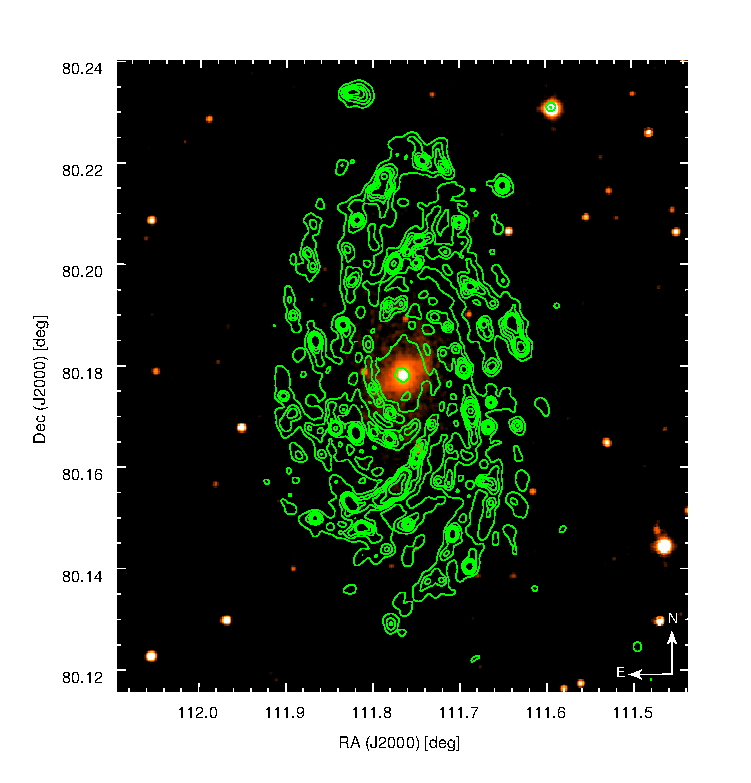}
\caption{ Overlay of UV images on 2MASS K band image. Contour map of NUV in Silica filter at the top and FUV image in BaF2 filter at the bottom. The contour levels are as in Figure~\ref{fig:uv_R}.}
\label{fig:uv_k}
\end{figure*}
\begin{figure*}
\centering
\includegraphics[scale=0.4]
{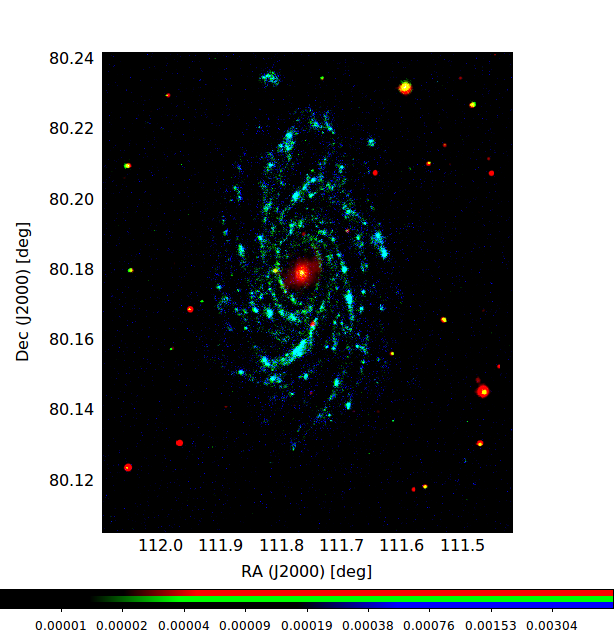}
\caption{ Three-color composite image of 2MASS-K band (red), UVIT NUVF1 (Green) and FUVF2 (Blue) of NGC 2336, which clearly shows the NIR (red) from the bar and UV emission (blue and green) from the spiral arms.}
\label{fig:rgb}
\end{figure*}
\begin{figure*}
\centering
\includegraphics[scale=0.4]{NUVF1.pdf}
\includegraphics[scale=0.4]{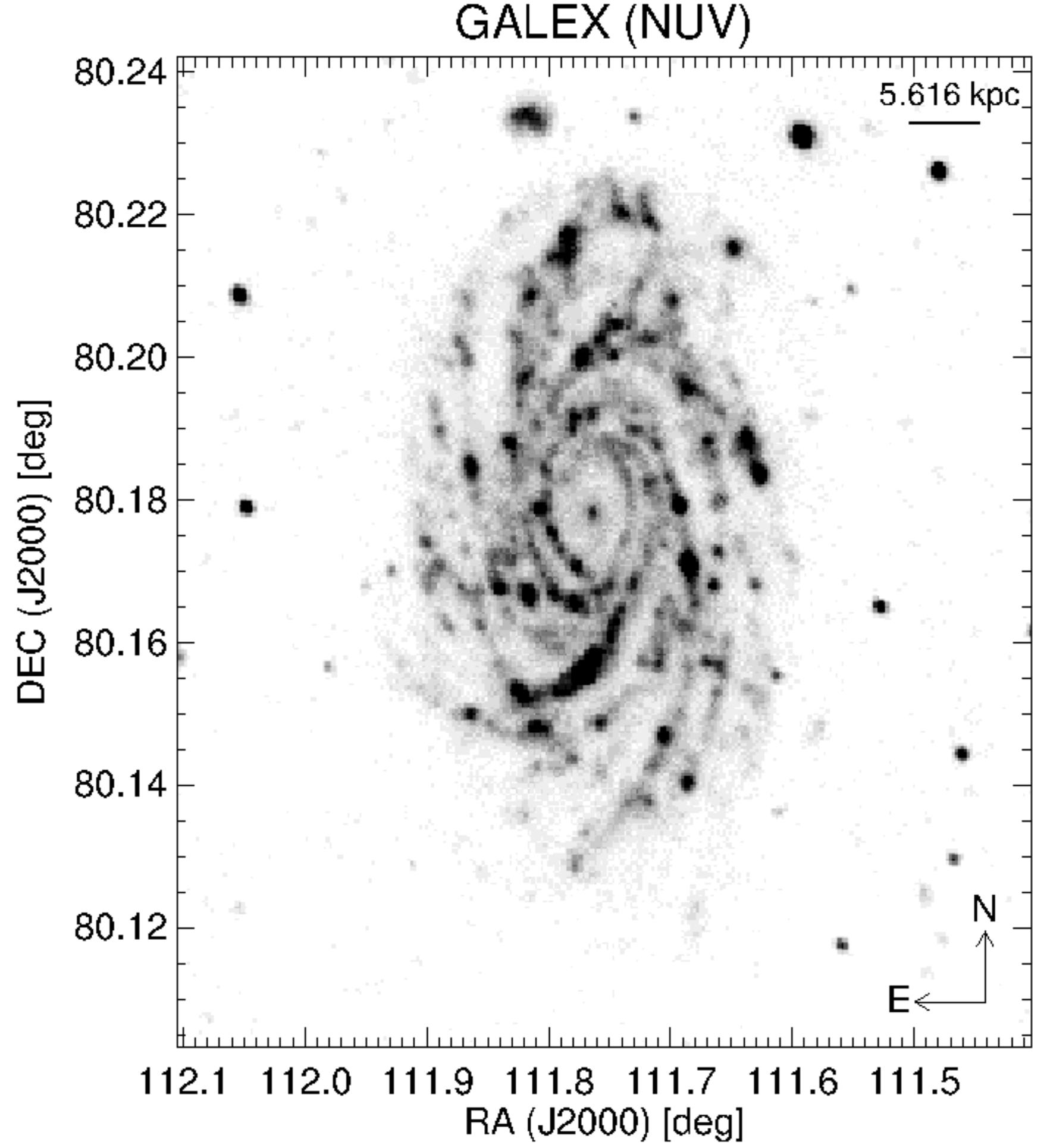}
\caption{Comparison of UVIT NUV image (Silica) with GALEX NUV image.}
\label{fig:nuvgalex}
\end{figure*}
\begin{figure*}
\centering
\includegraphics[scale=0.4]{FUVF2.pdf}
\includegraphics[scale=0.4]{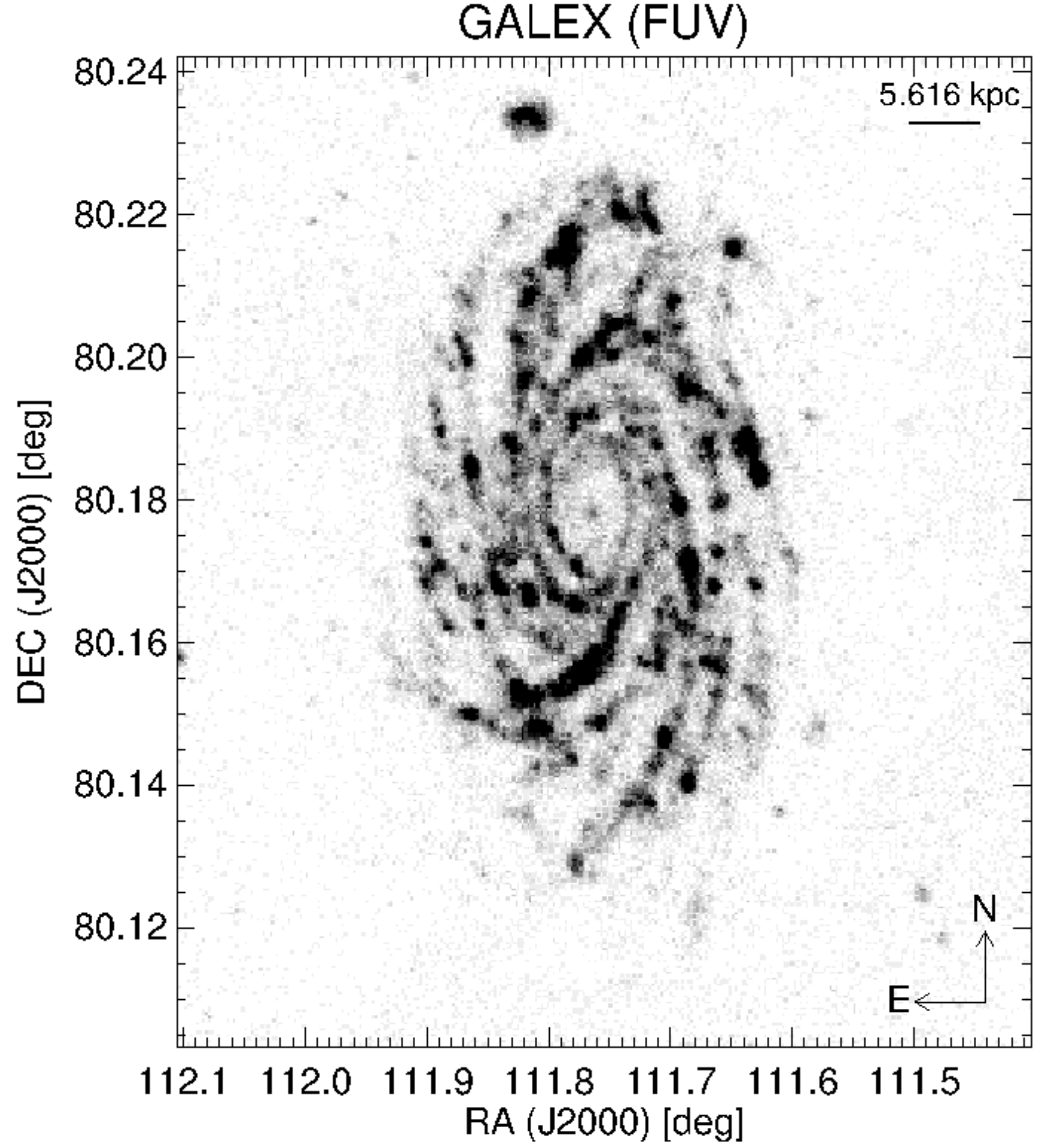}
\caption{Comparison of UVIT FUV (BaF2) images with GALEX FUV image.}
\label{fig:fuvgalex}
\end{figure*}

We have plotted contour maps of the UVIT-UV images of NGC 2336 on archival optical, H$\alpha$, and NIR images in Figures~\ref{fig:uv_R} to~\ref{fig:uv_Halpha}. The overlay of UVIT NUV and FUV contours on the R band image (Figure~\ref{fig:uv_R}) shows that the distribution of spiral arms in both images are same, but the nucleus, bulge, and bar are not clearly visible in UV. This is not surprising as the near IR emission is due to the presence of old stars in these regions. The star forming knots are clearly seen in UV images but not seen in the visual image. The nucleus in the UV image is not as bright in the optical image.

The morphological comparison between UVIT images and H$\alpha$ images shows several similar features in this galaxy. The ring around the bar is clearly identified in the UV images and also outlined in the H$\alpha$ images (Figure~\ref{fig:uv_Halpha}). Although, UV emission traces more evolved stars than H$\alpha$, there is a direct correlation between the spiral features in the UV and H$\alpha$ images.

 The morphological comparison of the 2MASS K band image with UVIT images are shown in Figure~\ref{fig:uv_k}. The spiral arms are not clearly seen in 2MASS K image. The bar is clearly seen in the K band indicating that the bar is composed of an older stellar population compared to the UV bright the spiral arms (Figure~\ref{fig:rgb}). However, there is no significant UV emission associated with bar ends. This suggests that the bar is not driving the disk star formation. Rather, it is driven by the multi-armed spiral structure.
 
We also compared UVIT images with GALEX images. The GALEX FOV is circular with a 1.2$^{\circ}$ diameter. It has two channels FUV and NUV with an image resolution of 4.3$^{\prime\prime}$ and 5.3$^{\prime\prime}$ respectively. The higher resolution of UVIT images reveals a clearer picture of multi armed spiral structure and the star forming regions associated with it compared to those obtained from GALEX (Figures~\ref{fig:nuvgalex} and~\ref{fig:fuvgalex}). 
%%%%%%%%%%%%%%%%%%%%%%%%%%%%%%%%%%%%%%%%%%%%%%%%%%%%%%%%%
\subsection {Identification of star-forming knots over the disk}
\begin{figure*}
\centering
\includegraphics[scale=0.5]{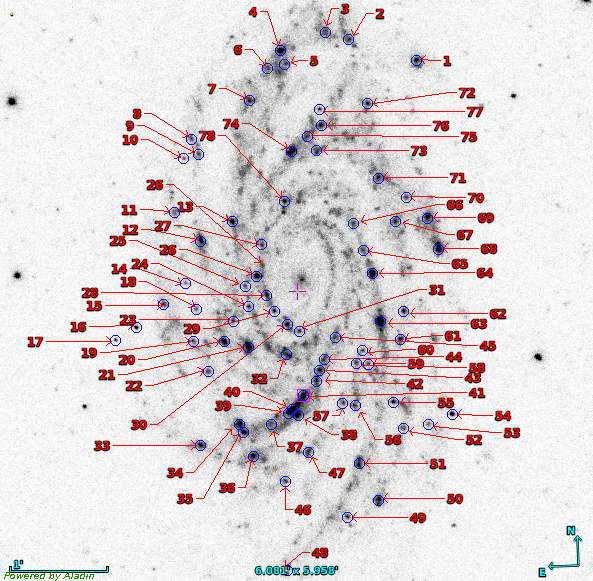}
\caption{star forming knotes in NUV image of UVIT.}
\label{fig:knots}
\end{figure*}

\begin{figure*}
\centering
\includegraphics[scale=0.55]{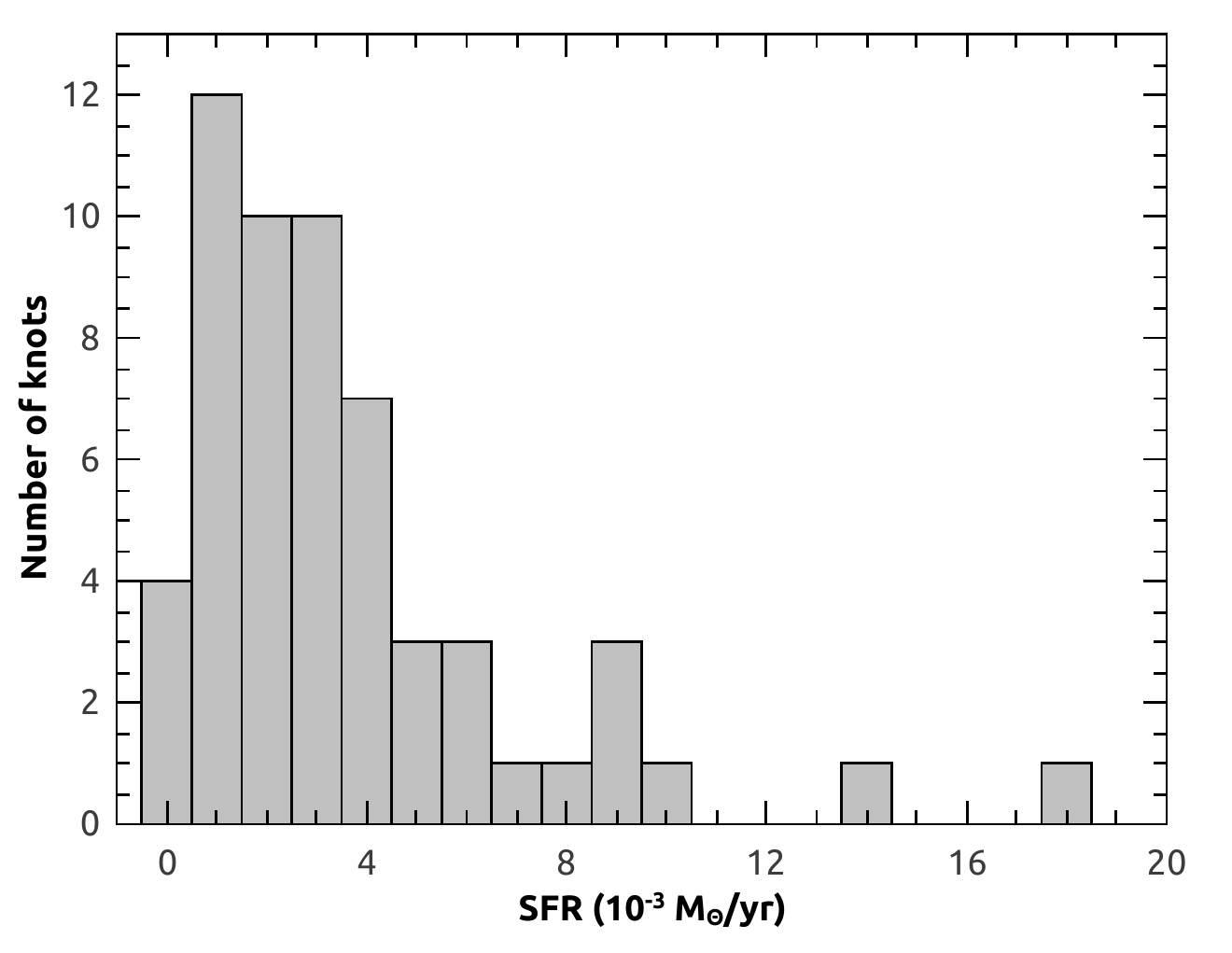}
\includegraphics[scale=0.55]{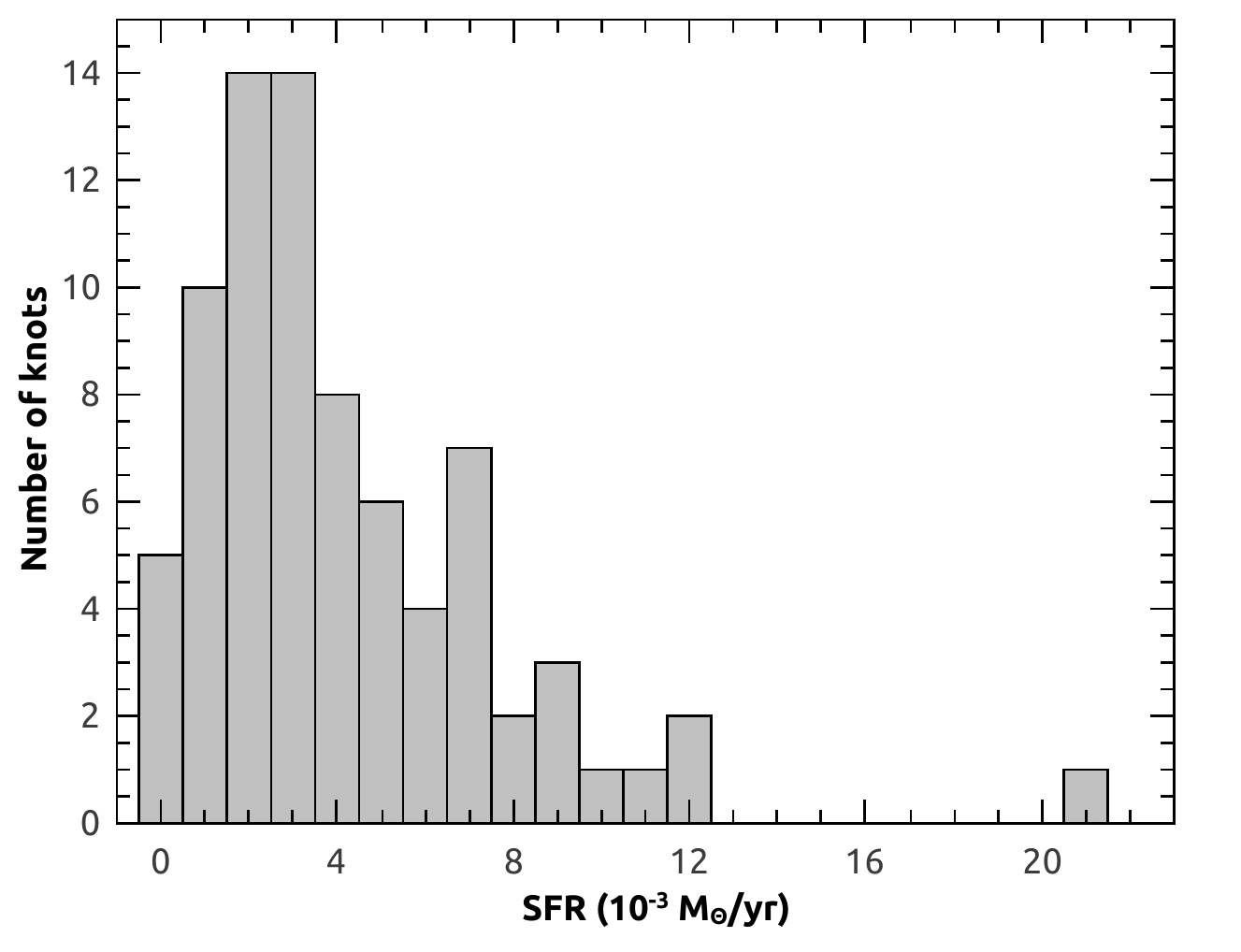}
\caption{Distribution of Star formation rate of knots in FUV (Left) and NUV (Right). Note that the SFR is in the units of 10$^{-3}$M$\odot$/yr.}
\label{fig:histsfr_knots}
\end{figure*}
\begin{figure*}
\centering
\includegraphics[scale=0.3]{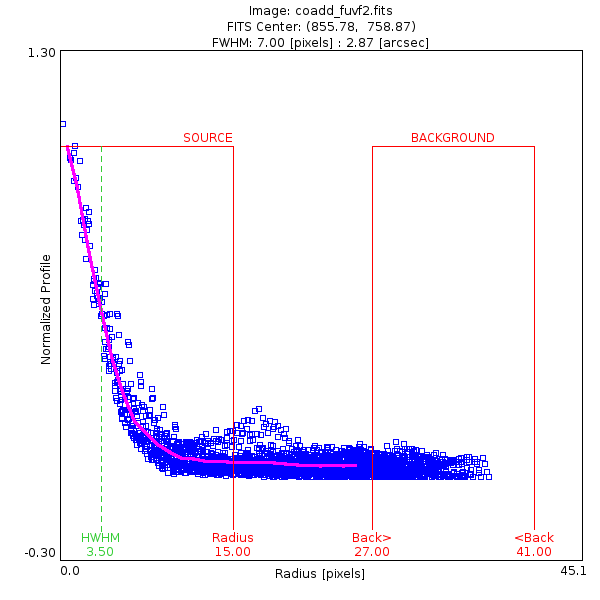}
\includegraphics[scale=0.3
]{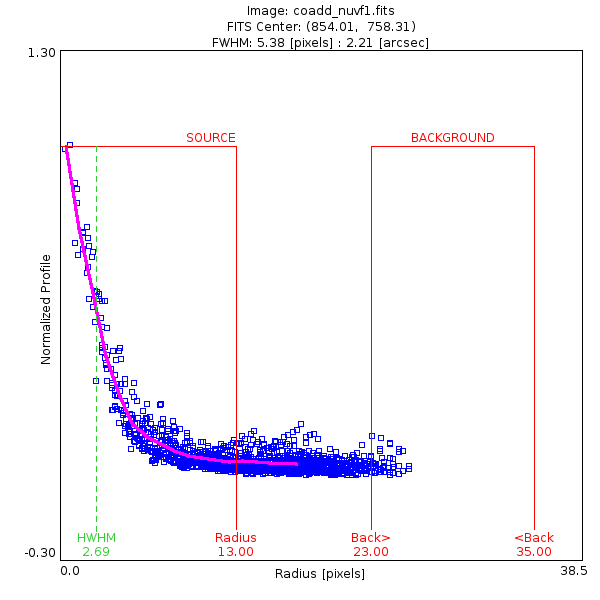}
\caption{Radial profile of a star forming knot (ID:64) in FUV (Left) and NUV (Right).}
\label{fig:radprof_sfknot}
\end{figure*}
\begin{figure*}
\centering
\includegraphics[scale=0.4]{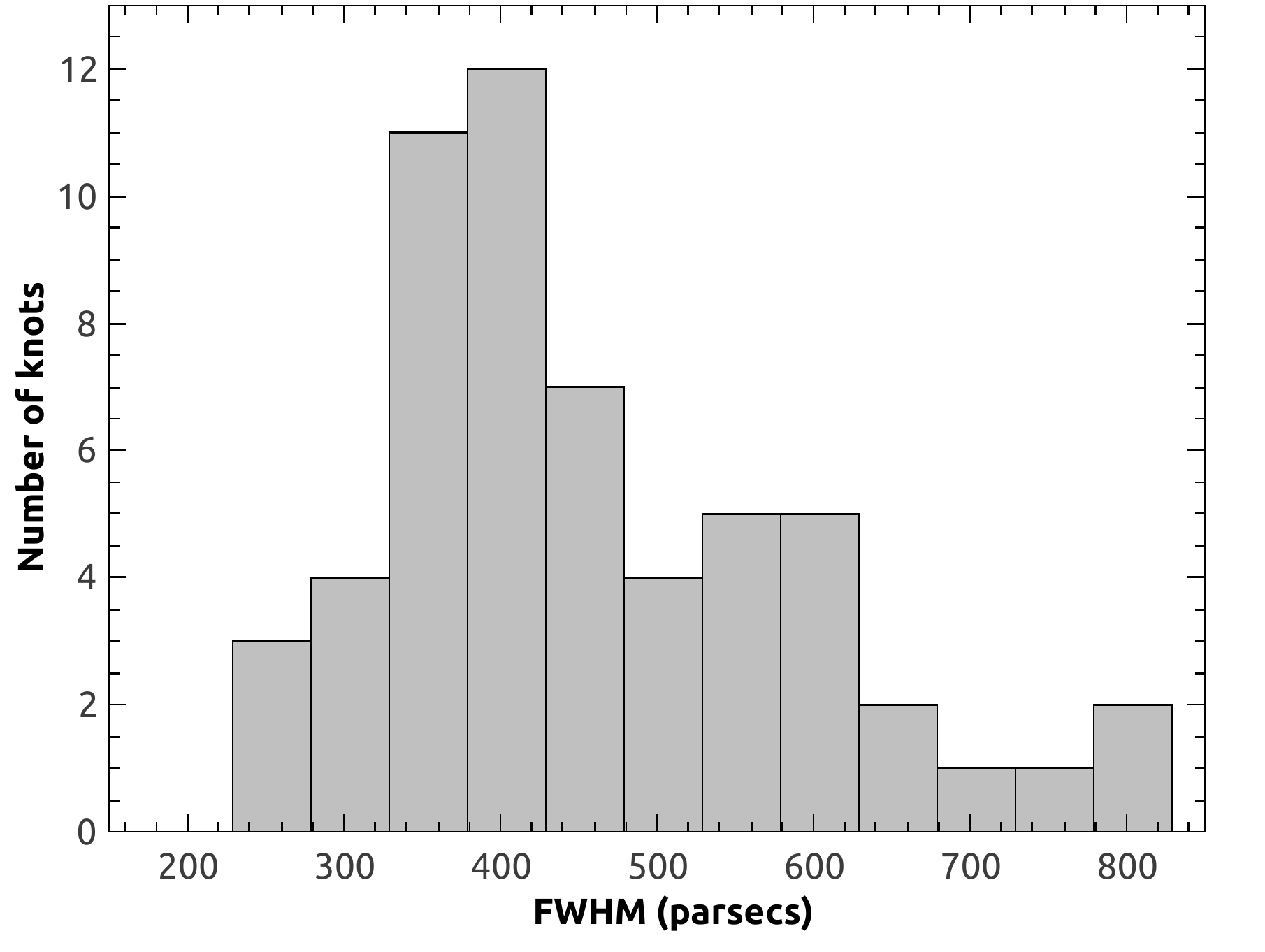}
\includegraphics[scale=0.4]{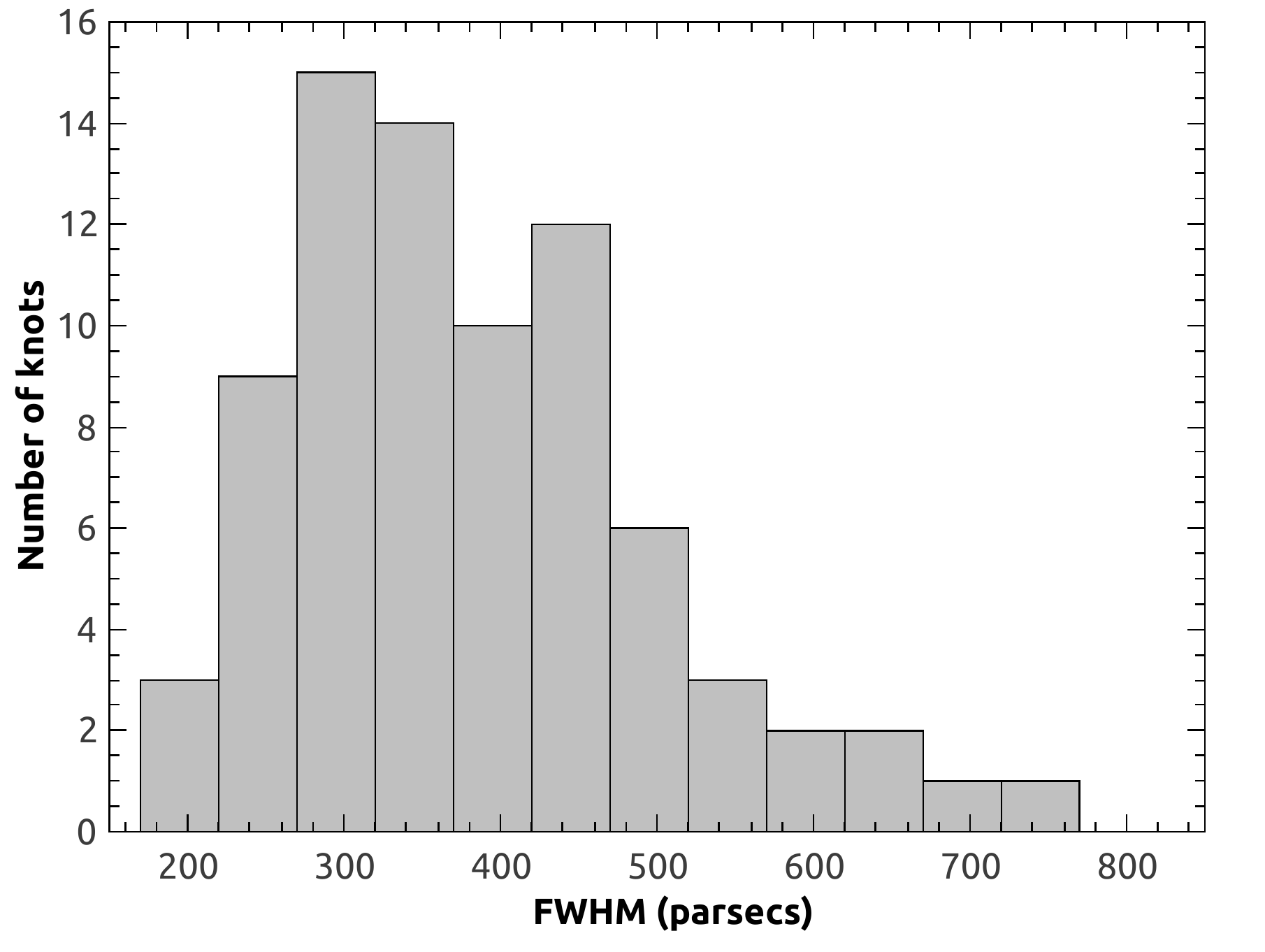}
\caption{Distribution of knot sizes for NGC 2336 in FUV (Left) and NUV (Right) UVIT images.}
\label{fig:size_knots}
\end{figure*}
\begin{figure*}
\centering
\includegraphics[scale=0.6]{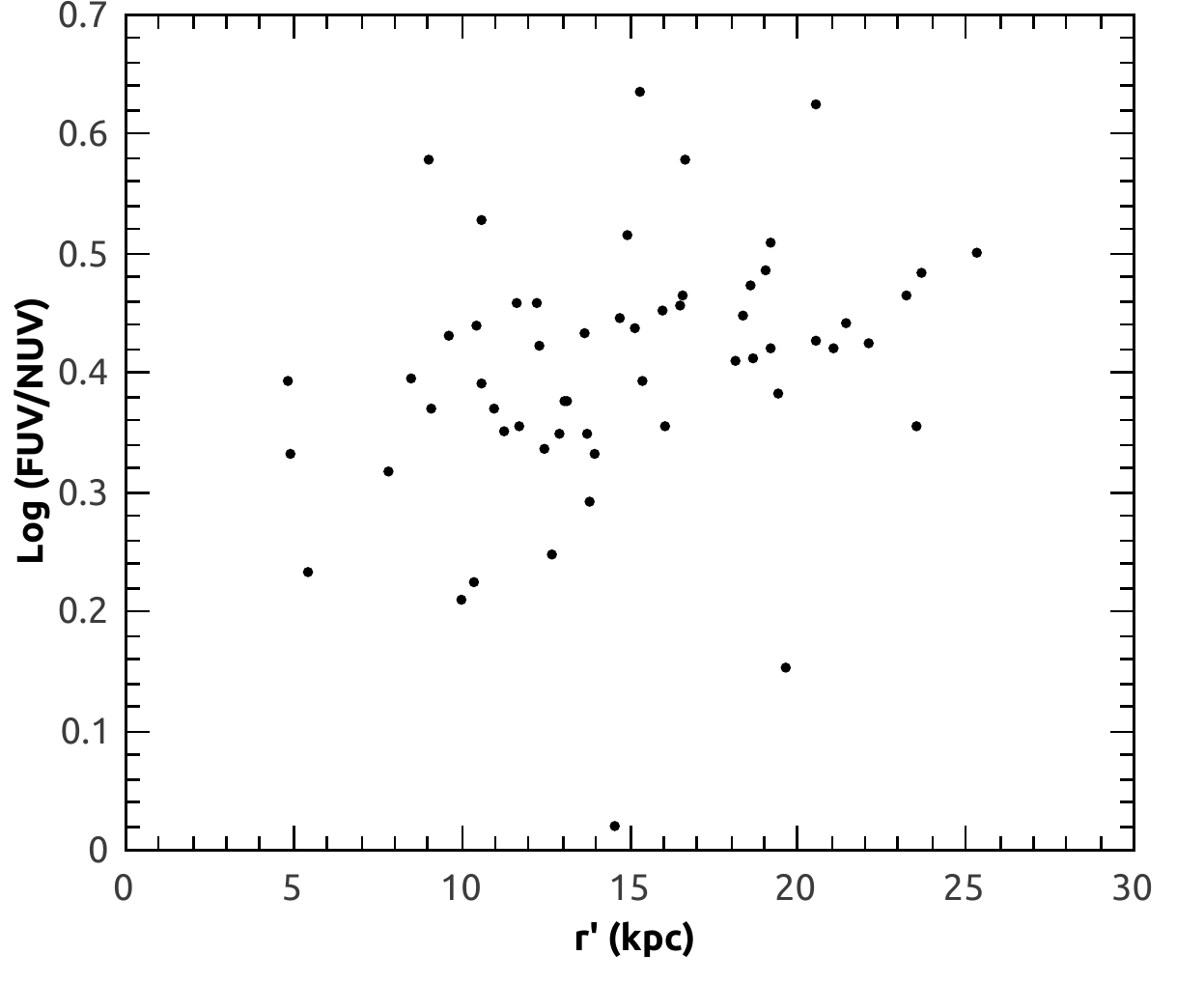}
\caption{Distribution of UV color with the deprojected distance (r') from the center of the galaxy.}
\label{fig:uvcolor}
\end{figure*}
\begin{figure*}
\centering
\includegraphics[scale=0.65]{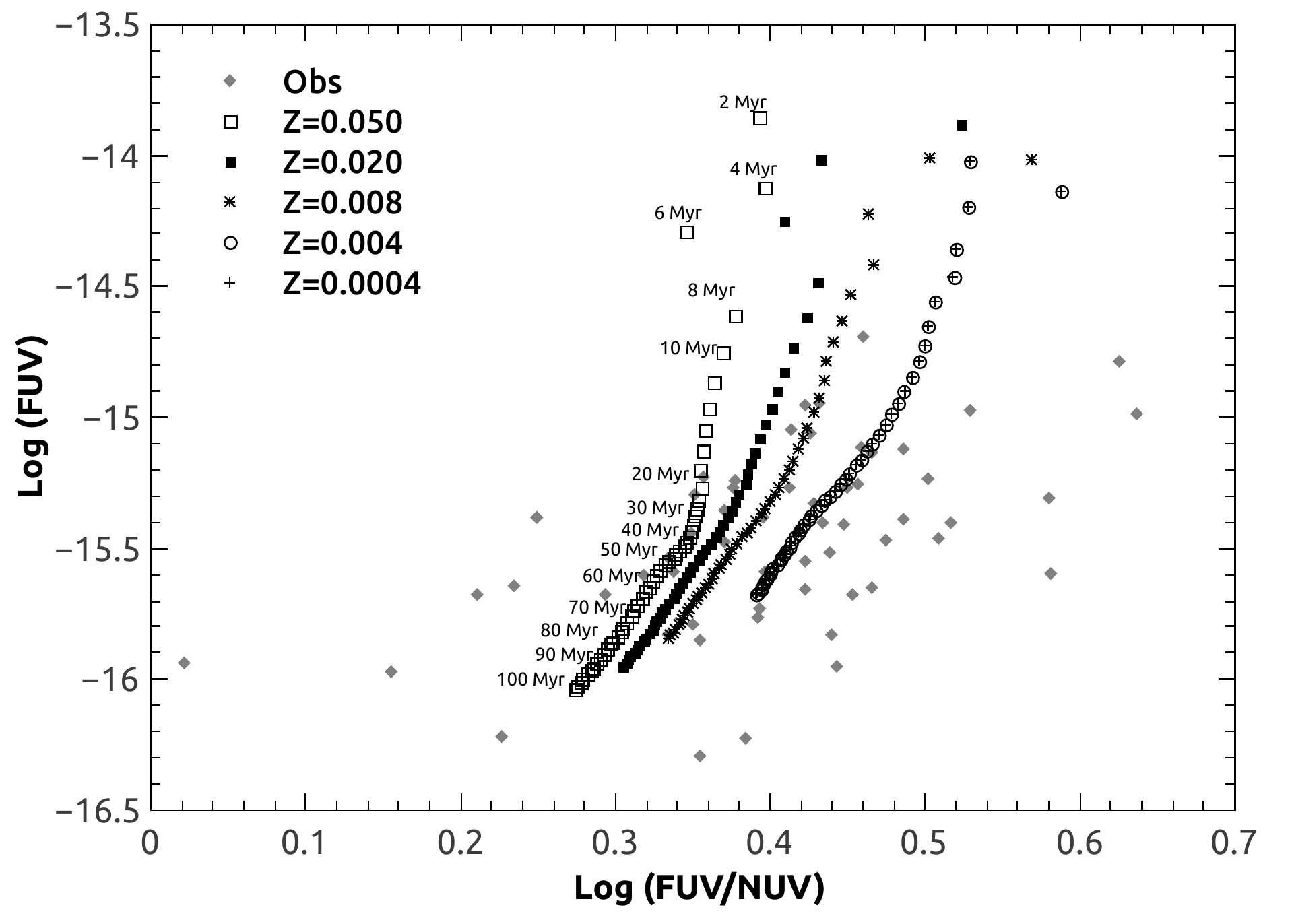}
\caption{Log (FUV) vs. UV color plot for star forming knots with the evolutionary track of Leitherer et al. (1999) in different metalicities.}
\label{fig:age}
\end{figure*}
Many spiral galaxies have bright knots along their spiral arms indicating that they are very young stellar clusters and their formation is triggered by a front associated with a density wave \citep{nikola2001}. These knots are star formation complexes containing multiple star clusters, rich in gas and dust \citep[e.g.][]{elmegreen2006, peterson2009star, smith2014extra}. The typical size of these clusters is $\le$ 5 pc, ages ranging from 1 to 10 Myr and luminosity functions can be fitted with power laws with exponents of approximately -2 \citep{whitmore2003}. Several studies have reported that very bright, young, star-forming regions are viable precursors of globular clusters \citep[e.g.][]{whitmore1993, schweizer1996}. A detailed study of such young stellar clusters can be used to study their star formation history and understand how density waves trigger the formation of massive stellar clusters \citep{grosbol2006bright}.
Based on the integrated fluxes and colors, UV knots have ages in the range of 0 - 200 Myr \citep{thilker2005recent, de2007galex, dong2008spitzer}.

NGC 2336 has numerous well developed, branching spiral arms and well-defined star-forming knots in the spiral arms. We have investigated the distribution of the knots of star formation over the disk of this spiral galaxy using those broadband filters with the best signal-to-noise ratio (NUV-Silica and FUV-BaF2).

We used an automated routine, the IDL (Interactive Data Language\footnote{http://www.harrisgeospatial.com/ProductsandTechnology/ \linebreak Software/IDL.aspx}) library routine {\it find.pro} (adapted from DAOPHOT: \citet{Stetson1987})  with FWHM of 1.2 pixels and sharplo, sharphi, roundlo, and roundhi set at values of 0.1, 1.2, -0.2 and 2.0 respectively.  This will allow the routine to pick slightly extended or/and elongated knots as recommended by \citet{smith2016}. After removing duplicates and checking the individual profiles for consistency, we identified 78 knots in the NUV and 57 knots in the FUV images (Figure~\ref{fig:knots}). Some of the NUV knots are not visible in FUV (Table~\ref{Tab:sfknots}). The distribution of these knots shows that they are clearly linked to the spiral arms as nearly all of them are associated with the spiral arms. The knots are concentrated more in the southern part of NGC 2336. We grouped these knots into loose knots and tight knots based on their radial profile. Among 78 knots, 30 knots are loose clusters and 48 knots are tight clusters.

We classified the knots into three basic groups: knots in the disk of the galaxy, knots in the co-rotation ring and knots in the galactic nuclei. We have identified 72 knots in the disk and 6 knots in the co-rotation ring in the NUV image. In FUV there are 54 knots in the disk and 3 knots in the co-rotation ring. Most of the knots are distributed over the spiral arms of the disk. Only 2 knots are in the inter arm region.

We performed aperture photometry of the knots and found that the luminosity is higher in the FUV, indicating recent star formation in the spiral arms (Table~\ref{Tab:sfknots}). We estimated the SFR of the knots using the relation from \cite{calzetti2013star}:
\begin{equation}
SFR_{UV}=C \times L_{UV} \times \lambda_{UV},
\label{eq:sfr}
 \end{equation}
 where C is a constant, $L_{UV}$ is the luminosity of UV and $\lambda_{UV}$ is the wavelength of UV. The value of C is 3 $\times 10^{-47} $for FUV and 4.2 $\times 10^{-47} $for NUV.   Figure~\ref{fig:histsfr_knots} shows the histogram of the SFR of the knots in the FUV and NUV bands. The distribution shows a broadly peaked FUV and a sharp peak in the NUV. The average SFR of the complete sample of star forming knots is 4.7 $\times 10^{-3} M_{\odot}/yr$ in NUV and 4.1 $\times 10^{-3} M_{\odot}/yr$ in FUV. The knot with the highest SFR is located in the northern arm at a distance of 11.6 kpc from the galaxy center(id no. 63).
 
The size of each knot is determined from the FWHMs of the knots by fitting Gaussians to the knots (Figure~\ref{fig:radprof_sfknot}). Figures~\ref{fig:size_knots} shows the size distribution of the knots, which is peaked at 425.9 pc in FUV and 344.8 pc in NUV. Most of the knots have an FWHM of approximately 300 to 450 parsecs. Our results agree with the results obtained by \cite{gusev2003structure}. The median size of the loose clusters in NUV is 469.6 pc and in FUV is 510.1 pc. The median size of the tight clusters in NUV is 355.7 pc and in FUV is 425.9 pc. The largest star forming knot in NUV (id no. 40) is associated with the northern major spiral arm and is at a distance of 12.6 kpc from the center in the southern part of galaxy. The largest star forming knot in FUV (id no. 3) is associated with the southern major spiral arm is at a distance of 23.8 kpc from the center in the northern part of galaxy. Some of the knots may have FUV emission from the surrounding molecular gas (H$_{2}$) which remains after star formation. This could be the reason why some of the knots are brighter in FUV and have larger sizes than in NUV.

We calculated the distance (r) and the de-projected distance ($r^{\prime}$) from
the center of NGC 2336 to the star formation region (Table~\ref{Tab:sfknots2}). The de-projected distance ($r^{\prime}$) is calculated using Eq. \ref{eq:real distance}, 
\begin{equation}
deprojected\:distance,{r}'=\frac{r}{\sqrt{\cos^2(\theta )+\sin^2(\theta )\cos^2(i)}},
\label{eq:real distance}
 \end{equation}
\begin{equation}
\theta=cos^{-1}(\frac{\delta DEC}{ \sqrt{(\delta RA)^2+(\delta DEC)^2}}), 
\label{eq:theta}
 \end{equation}
where r is the observed distance from the center of the galaxy, i is the inclination and $\theta$ is calculated from Eq. \ref{eq:theta} which is the angle between the major axis and radial distance to the knot \citep{martin1995, gadotti2007, honey2016}. 

The UV color (FUV-NUV) is calculated and plotted with $r^{\prime}$ in Figure~\ref{fig:uvcolor}. The knots that were not identified in both FUV and NUV images were removed before calculating the UV colors. The spatial distribution of UV color in the knots clearly appears bluer in the center and redder as the distance from the center increases, suggesting that star formation is stronger in the central regions \citep{Hicks2010}. The distribution in Figure~\ref{fig:uvcolor} also shows that the most of the knots in the ring are in the bluest region. The brightest star forming knot in both NUV and FUV (id no. 63) is associated with the northern major spiral arm and is at a distance of 11.6 kpc from the center in the western part of galaxy.

The UV colors can be used to estimate the age of the stellar population in the galaxy \citep{Bianchi2011}. We have used \cite{leitherer1999} evolutionary models to estimate ages of star-forming knots in NGC 2336. We ran Starburst99 simulations to obtain the evolution models from \cite{leitherer1999}. We set the parameters by assuming instantaneous star formation, Kroupa IMF ($\gamma=1.3, 2.3$), the total mass of stars assumed is $10^{6} M_{\odot}$ with mass range 1- 100 $M_{\odot}$, and range of age of the models is $10^{6}-10^{9}$ yr. We also ran Starburst99 with Salpeter IMF and different evolutionary tracks. We found that the result from Padova with IMF of Kroupa match well with our data. We used all possible values of metallicity (Z=0.0004, 0.004, 0.008, 0.020 and 0.050, where Z=0.020 equals solar abundance) in Padova to run the simulation. The model does not include nebular emission lines. We convolved the model spectrum with the UVIT filter curves to get the model UV fluxes. Figure~\ref{fig:age} shows the distribution of observed colors of the star forming knots with the evolutionary tracks in different metallicities. The observed color of the knots is compared with the model values and best age of the knot defined by taking the age of the closest model data point. The average value of estimated age is 57.3 Myr. Most of the knots in the complete sample have age $\le$ 70 Myr (68\%). Figure~\ref{fig:age} shows that fifteen knots are close to the evolutionary track with high metallicity of 2.5 Z$\odot$, five knots are close to the evolutionary track with solar metallicity Z$\odot$, thirteen knots are close to the evolutionary track with metallicity 0.4 Z$\odot$, eleven knots are close to the evolutionary track with metallicity 0.2 Z$\odot$, and thirteen knots are close to the evolutionary track with low metallicity 0.02 Z$\odot$. We observed that there is no obvious correlation between age and position.

To estimate the masses of the knots, we have followed the method used in \cite{hancock2003}, where they have determined the mass from the ratio of the observed FUV flux of the knot to the model FUV flux  at the point where the best age is determined and multiplied with the mass of the model ($10^{6} M_{\odot}$). The average value of the estimated mass of the knot is 9.8 $\times 10^{5} M_{\odot}$ and the median value of the total mass of the knots is 9.9 $\times 10^{5} M_{\odot}$. This result is similar to the average total mass of star-forming knots in NGC 3396 \citep{hancock2003}. The properties of the co-rotation ring is explained in the next section. 
%%%%%%%%%%%%%%%%%%%%%%%%%%%%%%%%%%%%%%%%%%%%%%%%%%%%%%%%%
\subsection {UV emission from the co-rotation ring}
\begin{figure*}
\centering
\includegraphics[scale=0.4]
{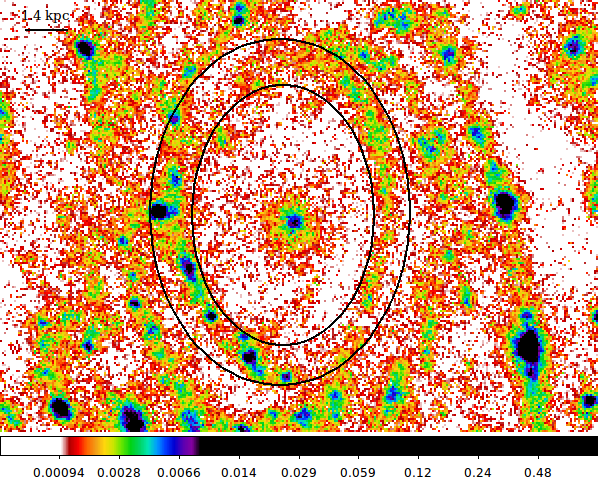}
\caption{ Nuclear ring of NGC 2336.}
\label{fig:nf1ring}
\end{figure*}
\begin{table*}
\centering
\caption{Photometry of co-rotation ring of NGC 2336}
\label{tab:phot_ring}
%\resizebox{\textwidth}{!}{
\begin{tabular}{lcccccccc}
\toprule
  & \textbf{NUVF1} & \textbf{NUVF2} & \textbf{NUVF3} & \textbf{NUVF5} & \textbf{FUVF1} & \textbf{FUVF2} & \textbf{FUVF3} & \textbf{FUVF5} \\ 
  \midrule
CPS & 15.488 & 0.564 & 4.116 & 4.336 & 1.336 & 1.043 & 0.781 & 0.462 \\
CPS err & 0.073 & 0.013 & 0.029 & 0.026 & 0.027 & 0.016 & 0.014 & 0.010 \\
Flux $^{a}$ & 3.75E-15 & 4.10E-15 & 5.29E-15 & 3.76E-15 & 6.46E-15 & 5.48E-15 & 5.30E-15 & 8.24E-15 \\
Flux err $^{a}$ & 1.76E-17 & 9.44E-17 & 3.77E-17 & 2.25E-17 & 1.29E-16 & 8.30E-17 & 9.37E-17 & 1.75E-16 \\
surface brightness $^{b}$ & 1.09E-22 & 1.19E-22 & 1.54E-22 & 1.10E-22 & 1.88E-22 & 1.60E-22 & 1.55E-22 & 2.40E-22 \\
surface brightness err $^{b}$& 5.13E-25 & 2.75E-24 & 1.10E-24 & 6.57E-25 & 3.77E-24 & 2.42E-24 & 2.73E-24 & 5.11E-24 \\
Relative contribution $^{c}$& 3.8\% & 3.4\% & 3.8\% & 3.9\% & 2.8\% & 2.7\% & 2.8\% & 3.1\% \\
\bottomrule
 \multicolumn{9}{l}{$^{a}$ erg/s/cm$^{2}$/\AA} \\
\multicolumn{9}{l}{$^{b}$ erg/s/cm$^{2}$/\AA /pc$^{2}$} \\
\multicolumn{9}{l}{$^{c}$ Relative contribution is calculated as a ratio of ring flux to total galaxy flux}
\end{tabular}
%}
\end{table*}
The co-rotation and nuclear rings play an important role in the secular evolution of galaxies \citep{kormendy2004secular}. The nuclear rings in barred galaxies are sites of intense massive star formation and are produced by the infall of gas from large radii caused by the bar \citep{Combes1985, shlosman, athanassoula1992, heller1994, knapen1995, buta1996galactic}. The angular radius of the ring is 5.3 kpc (33.97 $^{\prime\prime}$) in B band \citep{gusev2003structure}, which is approximately equal to the bar semi-major axis \citep{boroson1981,gusev2003structure}. The angular radius in the  bar is 35$^{ \prime\prime}-37^{ \prime\prime}$ \citep{boroson1981,Wilke1999}. The radius of the star formation ring coincides with that of the molecular gas (H$_{2}$) ring (at 13.6 kpc) in the galaxy which indicates that star-forming regions are concentrated in the molecular ring of NGC 2336 \citep{gusev2003structure}. In this section, we investigate the star formation regions along the co-rotation ring in NGC 2336.

The UVIT images clearly show a ring of star-forming regions around the bar. We estimated the size of the ring by drawing two elliptical regions for inner and the outer rings (Figure~\ref{fig:nf1ring}). For outer ring, the semi-minor axis, a $=$ 26.7$^{\prime\prime}$ (4.2 kpc) and semi-major axis, b $=$ 36.1$^{\prime\prime}$ (5.6 kpc), For the inner ring, a $=$ 19.1$^{\prime\prime}$ (2.9 kpc) and b $=$ 26.9$^{\prime\prime}$ (4.2 kpc). The approximate UV width of the ring is 8.6$^{\prime\prime}$ (1.3 kpc). We calculated the flux within the co-rotation ring around the bar by subtracting the total counts of the inner elliptical region from the outer elliptical region in different filters (Table~\ref{tab:phot_ring}). The co-rotation ring contributes 3 to 4\% of the total UV flux of the galaxy. We identified 6 prominent star forming knots (Id: 26, 27, 28, 29, 30, and 31) in NUV and 3 star forming knots (Id: 28, 29, 30) in FUV in the east-south direction of the ring (Table~\ref{Tab:sfknots} and~\ref{Tab:sfknots2}). We could not find any star forming knots in the north-west direction. The SFR of these knots is given in Table~\ref{Tab:sfknots} and~\ref{Tab:sfknots2}. Among 6 star forming knots in the ring, id no. 26 has a higher SFR with the FWHM of 1.6$^{\prime\prime}$ (0.24 kpc) at a distance of 7 kpc. The star forming knots are not clustered at the bar ends but instead are distributed over one side of the co-rotation ring. Several well-developed spiral arms spring from this star-forming ring. In addition to the arms and co-rotation ring, we measured the SFR in the nucleus as explained in the next section.
%%%%%%%%%%%%%%%%%%%%%%%%%%%%%%%%%%%%%%%%%%%%%%%%%%%%%%%%%
\subsection {Nuclear emission}
\begin{figure*}
\centering
\includegraphics[scale=0.4]
{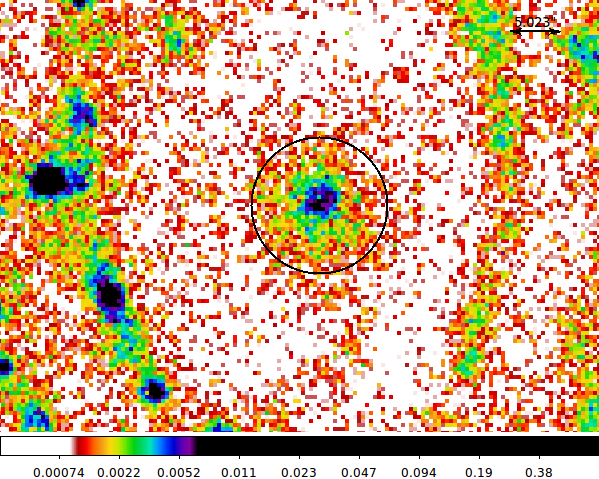}

\caption{Nucleus of NGC 2336.}
\label{fig:nf1nucls}
\end{figure*}

\begin{figure*}
\centering
\includegraphics[scale=0.3]
{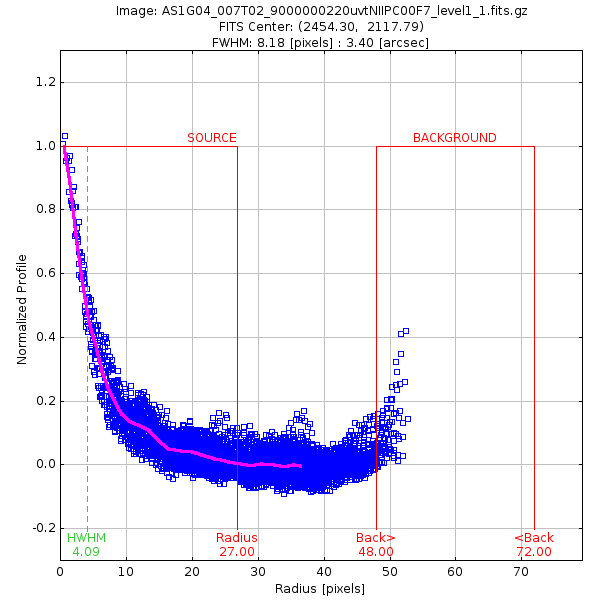}

\caption{Radial profile of nucleus of NGC 2336.}
\label{fig:nf1nucls_profile}
\end{figure*}

Figure~\ref{fig:nf1nucls} shows the UV emission from the nucleus of NGC 2336, which is not as bright in UV as the spiral arms. This galaxy hosts a Seyfert 2 type AGN. To explore the possible relationship between the AGN activity and massive star formation in the galaxy, we did aperture photometry of the central nuclear region of NGC 2336 in multiple UVIT observations. The aperture size of 7$^{\prime\prime}$ (1.1 kpc) was used for aperture photometry. Figure~\ref{fig:nf1nucls_profile} shows the radial profile of the nuclear region in the smoothed NUV image. The contribution of UV flux is 0.5 \% in NUV and 0.2\% in FUV to the total UV flux of the galaxy. The luminosity of the nuclear region is 1.5 $\times 10^{41}$ erg/s in NUV and 7.5$\times 10^{40}$ erg/s in FUV. The SFR of the nuclear region is 1.5~$\times 10^{-2} M_{\odot}/yr$ and 3.5 $\times 10^{-3} M_{\odot}/yr$ in NUV and FUV respectively.

\begin{figure*}
\centering
\includegraphics[scale=0.5]
{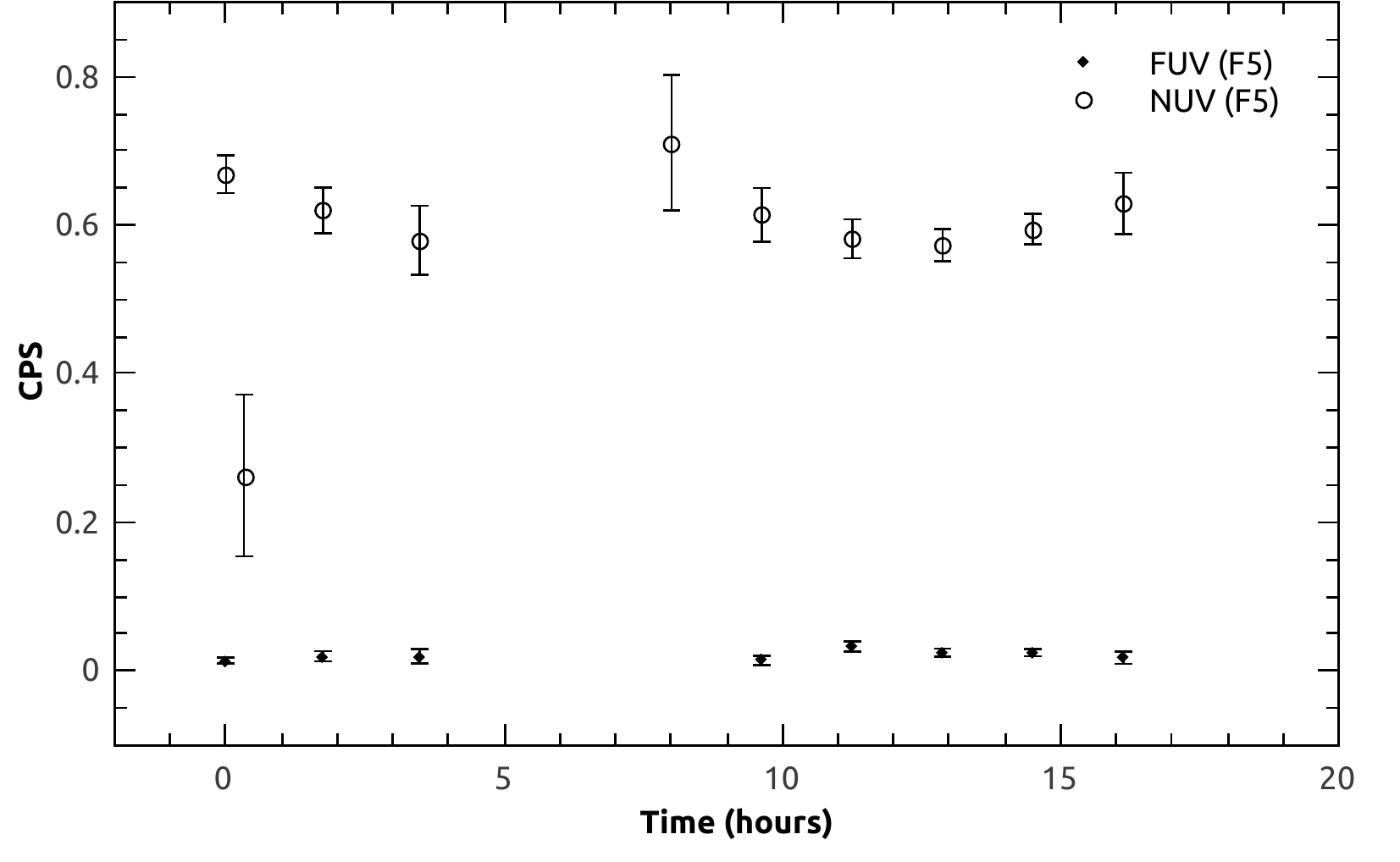}
\caption{UVIT count rate of nucleus as a function of time. open circle: NUV B4 filter. diamond: FUV silica.}
\label{fig:variability}
\end{figure*}

We have examined the variation of count rate over the total time of observation. We have used the narrow band filter F5 to check the variability. We could not find any significant variation in the count rate within 16 hrs of observations (Figure~\ref{fig:variability}). This indicates that the AGN does not shows any short-term variability in UV flux.  

%%%%%%%%%%%%%%%%%%%%%%%%%%%%%%%%%%%%%%%%%%%%%%%%%%%%%%%%%
 \subsection{Implications:}
The deep, high-resolution UVIT FUV and NUV observations of the nearly face-on spiral galaxy NGC~2336 provides an excellent opportunity to study the association and distribution of star-forming knots along the spiral arms in a disk galaxy. As UV emission traces massive stars (FUV) as well as less massive B and A-type stars (NUV), it can give us a better idea of how star formation is triggered by the spiral arms compared to only H$\alpha$ emission which traces only young O type stars that are about $10^{6}$ years old. The distribution of knot sizes and colors can also reveal how the star formation is associated with the spiral arms and spurs.

The knots are nearly always found along the arms and only 4\% are few found in the interarm regions. Therefore, star formation occurs nearly exclusively along the spiral arms as shown by earlier studies \cite[e.g.][]{calzetti2005}. However, the FUV-NUV color of the knots becomes redder at larger radii in the disk, indicating the presence of more long-lived star-forming knots at larger radii. This may be due to the lower tidal shear in the outer radii, which allows the clusters to live longer than those at smaller radii.

The FWHM of the knot sizes is different in the FUV compared to the NUV images. The knot FWHM sizes are sharply peaked at 426~pc in FUV and at 345~pc in NUV but vary uniformly between 300~pc to 450~pc in NUV emission. This may be due to the fact that FUV traces only massive stars that are short-lived compared to those traced by NUV emission. The median value of the FWHM of the knots in the NUV image is 389 pc and in FUV it is 448 pc. We compared the size of knots in this galaxy with the size of star-forming regions reported in \cite{Kennicutt1984}. They have reported the size of few well-known objects such as the diameter of the Orion nebula is 5 pc, Lagoon nebula(M8) is 25 pc, Rosetta nebula NGC 2244 is 50 pc, Carina complex (NGC3372) is 200 pc and 30 Doradus region in the Large Magellanic Cloud is ∼400 pc. The giant H II region NGC 5471 in the Pinwheel galaxy M101 has diameter of ∼800 pc and NGC 604 in Triangulum galaxy M33 has a diameter of ∼400 pc. Thus the knots in NGC 2336 are likely giant HII regions similar to the observed NGC 5471 and NGC 604 \citep{Kennicutt1984}.

The spiral arms are also thicker in NUV compared to FUV. This is again due to the wider distribution of masses traced by NUV emission. Thus NUV emission may be a better tracer of spiral structure in galaxies, especially those with less ongoing massive star formation. The average age of the star forming knots is 57 Myrs which suggests that these knots are older than the HII regions such as Rosetta, M16, RCW 79 and RCW 36 \citep{tremblin2014age}. The typical super star clusters have masses greater than $10^{5} M_{\odot}$ \cite{sternberg1998}. All the knots in NGC 2336 have masses greater than $10^{5} M_{\odot}$ suggesting that they have masses similar to the masses of super star clusters.
%%%%%%%%%%%%%%%%%%%%%%%%%%%%%%%%%%%%%%%%%%%%%%%%%%%%%%%%%
\section{Conclusions}
A summary of the main conclusions drawn from this study is discussed below.
\begin{enumerate}
\item We have studied the morphology of NGC 2336 in FUV and NUV using UVIT and compared it with the optical, H$\alpha$ and 2MASS K band images. The UV emission shows that there is widespread star formation over the galaxy but is most intense along the spiral arms and the corotation ring around the small, but distinct bar in the center of the galaxy.\\
\item  Using the high spatial resolution of UVIT we have identified individual star forming regions or knots in the disk of the galaxy. We have identified 78 individual knots in NUV and 57 knots in FUV. Among them, 6 knots in NUV and 3 knots in FUV are from the co-rotation ring; the remaining knots are associated with the spiral arms and its branches in the disk of the galaxy. Most of the knots are on the spiral arms; only 3 knots are found in the interarm regions of the spiral structure. There are 29 loose knots and 49 tight knots in NUV, 17 loose knots and 40 tight knots in FUV.\\
\item We measured the integrated FUV and NUV fluxes, luminosities, sizes, star formation rates (SFRs) and FUV-NUV color of the individual knots. We also estimated the ages and masses of the knots using \cite{leitherer1999} evolutionary models. The average luminosity of the complete sample of knots is 4.6$\times 10^{40}$ erg/s in NUV and 8.9$\times 10^{40}$ erg/s in FUV with the range of 6.7$\times 10^{39}$ erg/s to 2.1$\times 10^{41}$ erg/s in NUV and 9.8$\times 10^{39}$ to 3.9$\times 10^{41}$ erg/s in FUV. The average sizes of the star forming knots in NGC~2336 is 408~pc in NUV and 485~pc in FUV; the size ranges are (195--766)~pc in NUV and (254--810)~pc in FUV. The knots sizes are similar to that of giant HII regions NGC~5471 and NGC~604. The star formation in the knots varies from $6.8 \times 10^{-4}$ to $2.2 \times 10^{-2} M_{\odot}/yr$ in NUV and $4.5 \times 10^{-4}$ to $1.8 \times 10^{-2} M_{\odot}/yr$ in FUV with the average value of 4.7 $\times 10^{-3} M_{\odot}/yr$ in NUV and 4.1 $\times 10^{-3} M_{\odot}/yr$ in FUV respectively. There are a larger number of knots in the southern half of NGC~2336 which agrees with the earlier observations of \cite{gusev2003structure}.\\
\item The average value of the age of the star forming knots is 57.3 Myrs with the range of 14 to 100~Myrs. The comparison with models from \cite{leitherer1999} indicates that almost all the knots have the age of $\leq$ 70~Myr. The youngest of the knots is about 14~Myr old and has a mass of approximately 1.1 $\times 10^{6} M_{\odot}$. The average mass of star forming knots is 9.75 $\times 10^{5} M_{\odot}$ with the range of 5.57 $\times 10^{5}$  to 1.1 $\times 10^{6} M_{\odot}$. All the star-forming knots have masses  (>$10^{5} M_{\odot}$)  similar to the mass of typical super star clusters.\\
\item We have traced the knot color with radial distance from the center of NGC 2336. We find that the star forming knots have bluer (FUV-NUV) colors near the center and become increasingly redder with larger radii. This suggests that the star formation is more intense in the inner disk compared to larger radii in the galaxy. This is not surprising as the effect of the spiral arms is stronger and the orbital time scales are shorter near the galaxy center. The tidal shear is also greater near the center and so clusters live longer at larger radii.
\end{enumerate}
%%%%%%%%%%%%%%%%%%%%%%%%%%%%%%%%%%%%%%%%%%%%%%%%%%%%%%%%%
\section*{Acknowledgements}
This research has made use of NASA Astrophysics Data System Bibliographic Services. Many people at IIA, ISRO, IUCAA, TIFR, NRC (Canada) and Univ. of Calgary have contributed to different parts of the spacecraft, instrument and the operations. We also acknowledge the Gnu Data Language (GDL), the IDL Astronomy Library and its many contributors. This research has made use of the NASA/IPAC Extragalactic Database (NED) which is operated by the Jet Propulsion Laboratory, California Institute of Technology, under contract with the National Aeronautics and Space Administration. 

GALEX (Galaxy Evolution Explorer) is a NASA Small Explorer, launched in April 2003. We gratefully acknowledge NASA’s support for construction,  operation, and science analysis of the GALEX mission, developed in cooperation with the Centre National d’Etudes Spatiales of France and the Korean Ministry of Science and Technology. Some of the data presented in this paper were obtained from the Mikulski Archive for Space Telescopes (MAST). STScI is operated by the Association of Universities for Research in Astronomy, Inc., under NASA contract NAS5-26555. Support for MAST for non-HST data is provided by the NASA Office of Space Science via grant NNX09AF08G and by other grants and contracts.

This publication makes use of data products from the Two Micron All Sky Survey, which is a joint project of the University of Massachusetts and the Infrared Processing and Analysis Center/California Institute of Technology, funded by the National Aeronautics and Space Administration and the National Science Foundation.

The Digitized Sky Surveys were produced at the Space Telescope Science Institute under U.S. Government grant NAG W-2166. The images of these surveys are based on photographic data obtained using the Oschin Schmidt Telescope on Palomar Mountain and the UK Schmidt Telescope.
%%%%%%%%%%%%%%%%%%%%%%%%%%%%%%%%%%%%%%%%%%%%%%%%%%%%%%%%%
\appendix
\section{Properties of the star forming knots in NGC 2336}
\begin{landscape}
\begin{table}
\centering
\caption{Observed properties of the star forming knots}
\label{Tab:sfknots}
\resizebox{1.3\textwidth}{!}{
\begin{tabular}{cccccccccccccccc}
\toprule
\multirow{2}{*}{ID} & \multirow{2}{*}{\thead{RA\\ {[}Degrees{]}}} & \multirow{2}{*}{\thead{DEC\\ {[}Degrees{]}}} & \multicolumn{6}{c}{NUV} & \multicolumn{6}{c}{FUV} & \multirow{2}{*}{Tight/Loose}\\ \cline{4-15}
 &  &  & \multicolumn{1}{c}{CPS} & \multicolumn{1}{c}{err} & \multicolumn{1}{c}{\thead{Flux\\ (erg/s/cm2/A)}} & \multicolumn{1}{c}{\thead{err\\ (erg/s/cm2/A)}} & \multicolumn{1}{c}{\thead{Log($L_{NUV}$)\\(erg/s)}} & \multicolumn{1}{c}{\thead{err\\(erg/s)}} & \multicolumn{1}{c}{CPS} & \multicolumn{1}{c}{err} & \multicolumn{1}{c}{\thead{Flux\\ (erg/s/cm2/A)}} & \multicolumn{1}{c}{\thead{err\\ (erg/s/cm2/A)}} & \multicolumn{1}{c}{\thead{Log($L_{FUV}$)\\(erg/s)}} & \multicolumn{1}{c}{\thead{err\\(erg/s)}} &  \\ 
\midrule
1 & 111.6518 & 80.2157 & 0.757 & 0.016 & 1.83E-16 & 3.89E-18 & 5.50E+40 & 1.17E+39 & 0.111 & 0.005 & 5.83E-16 & 2.71E-17 & 1.12E+41 & 5.21E+39 & T \\
2 & 111.7200 & 80.2193 & 0.554 & 0.014 & 1.34E-16 & 3.33E-18 & 4.03E+40 & 9.98E+38 & 0.078 & 0.004 & 4.11E-16 & 2.28E-17 & 7.90E+40 & 4.37E+39 & T \\
3 & 111.7433 & 80.2204 & 1.006 & 0.018 & 2.44E-16 & 4.48E-18 & 7.31E+40 & 1.34E+39 & - & - & - & - & - & - & L \\
4 & 111.7887 & 80.2173 & 1.342 & 0.021 & 3.25E-16 & 5.18E-18 & 9.75E+40 & 1.55E+39 & 0.165 & 0.006 & 8.67E-16 & 3.30E-17 & 1.67E+41 & 6.35E+39 & T \\
5 & 111.7843 & 80.2151 & 0.533 & 0.013 & 1.29E-16 & 3.26E-18 & 3.87E+40 & 9.79E+38 & - & - & - & - & - & - & L \\
6 & 111.8013 & 80.2143 & 0.893 & 0.017 & 2.16E-16 & 4.22E-18 & 6.49E+40 & 1.27E+39 & - & - & - & - & - & - & L \\
7 & 111.8199 & 80.2088 & 0.785 & 0.016 & 1.90E-16 & 3.96E-18 & 5.71E+40 & 1.19E+39 & 0.102 & 0.005 & 5.36E-16 & 2.60E-17 & 1.03E+41 & 4.99E+39 & T \\
8 & 111.8781 & 80.2022 & 0.377 & 0.011 & 9.14E-17 & 2.74E-18 & 2.74E+40 & 8.23E+38 & - & - & - & - & - & - & L \\
9 & 111.8708 & 80.1996 & 0.392 & 0.012 & 9.50E-17 & 2.80E-18 & 2.85E+40 & 8.39E+38 & - & - & - & - & - & - & T \\
10 & 111.8854 & 80.1989 & 0.102 & 0.006 & 2.47E-17 & 1.43E-18 & 7.41E+39 & 4.28E+38 & 0.011 & 0.002 & 5.99E-17 & 8.68E-18 & 1.15E+40 & 1.67E+39 & T \\
11 & 111.8944 & 80.1897 & 0.444 & 0.012 & 1.08E-16 & 2.98E-18 & 3.23E+40 & 8.93E+38 & 0.066 & 0.004 & 3.48E-16 & 2.09E-17 & 6.68E+40 & 4.02E+39 & L \\
12 & 111.8684 & 80.1848 & 0.979 & 0.018 & 2.37E-16 & 4.42E-18 & 7.12E+40 & 1.33E+39 & 0.196 & 0.007 & 1.03E-15 & 3.60E-17 & 1.98E+41 & 6.92E+39 & T \\
13 & 111.8367 & 80.1881 & 0.771 & 0.016 & 1.87E-16 & 3.92E-18 & 5.60E+40 & 1.18E+39 & 0.083 & 0.004 & 4.38E-16 & 2.35E-17 & 8.43E+40 & 4.52E+39 & T \\
14 & 111.8834 & 80.1775 & 0.110 & 0.006 & 2.67E-17 & 1.48E-18 & 8.00E+39 & 4.45E+38 & - & - & - & - & - & - & T \\
15 & 111.9054 & 80.1739 & 0.439 & 0.012 & 1.06E-16 & 2.96E-18 & 3.19E+40 & 8.88E+38 & 0.054 & 0.004 & 2.82E-16 & 1.88E-17 & 5.41E+40 & 3.62E+39 & T \\
16 & 111.9328 & 80.1700 & 0.302 & 0.010 & 7.32E-17 & 2.46E-18 & 2.20E+40 & 7.37E+38 & - & - & - & - & - & - & T \\
17 & 111.9534 & 80.1678 & 0.148 & 0.007 & 3.59E-17 & 1.72E-18 & 1.08E+40 & 5.16E+38 & - & - & - & - & - & - & T \\
18 & 111.8727 & 80.1732 & 0.206 & 0.008 & 4.99E-17 & 2.03E-18 & 1.50E+40 & 6.08E+38 & - & - & - & - & - & - & L \\
19 & 111.8758 & 80.1676 & 0.217 & 0.009 & 5.26E-17 & 2.08E-18 & 1.58E+40 & 6.24E+38 & - & - & - & - & - & - & L \\
21 & 111.8450 & 80.1676 & 1.107 & 0.019 & 2.68E-16 & 4.70E-18 & 8.05E+40 & 1.41E+39 & 0.147 & 0.006 & 7.72E-16 & 3.12E-17 & 1.48E+41 & 5.99E+39 & T \\
21 & 111.8207 & 80.1667 & 1.731 & 0.024 & 4.20E-16 & 5.88E-18 & 1.26E+41 & 1.76E+39 & 0.216 & 0.007 & 1.13E-15 & 3.78E-17 & 2.18E+41 & 7.27E+39 & T \\
22 & 111.8608 & 80.1624 & 0.313 & 0.010 & 7.59E-17 & 2.50E-18 & 2.28E+40 & 7.50E+38 & 0.036 & 0.003 & 1.88E-16 & 1.54E-17 & 3.61E+40 & 2.96E+39 & T \\
23 & 111.8356 & 80.1711 & 0.220 & 0.009 & 5.33E-17 & 2.10E-18 & 1.60E+40 & 6.29E+38 & 0.028 & 0.003 & 1.47E-16 & 1.36E-17 & 2.82E+40 & 2.61E+39 & T \\
24 & 111.8201 & 80.1736 & 0.192 & 0.008 & 4.65E-17 & 1.96E-18 & 1.40E+40 & 5.87E+38 & - & - & - & - & - & - & T \\
25 & 111.8238 & 80.1771 & 0.111 & 0.006 & 2.69E-17 & 1.49E-18 & 8.07E+39 & 4.47E+38 & - & - & - & - & - & - & L \\
26 & 111.8121 & 80.1788 & 1.325 & 0.021 & 3.21E-16 & 5.14E-18 & 9.63E+40 & 1.54E+39 & - & - & - & - & - & - & T \\
27 & 111.8071 & 80.1842 & 0.346 & 0.011 & 8.39E-17 & 2.63E-18 & 2.52E+40 & 7.88E+38 & - & - & - & - & - & - & L \\
28 & 111.8021 & 80.1756 & 0.549 & 0.014 & 1.33E-16 & 3.31E-18 & 3.99E+40 & 9.93E+38 & 0.043 & 0.003 & 2.28E-16 & 1.69E-17 & 4.39E+40 & 3.26E+39 & T \\
29 & 111.7947 & 80.1729 & 0.284 & 0.010 & 6.88E-17 & 2.38E-18 & 2.06E+40 & 7.14E+38 & 0.028 & 0.003 & 1.49E-16 & 1.37E-17 & 2.86E+40 & 2.63E+39 & T \\
30 & 111.7816 & 80.1705 & 0.687 & 0.015 & 1.67E-16 & 3.70E-18 & 4.99E+40 & 1.11E+39 & 0.079 & 0.004 & 4.13E-16 & 2.28E-17 & 7.95E+40 & 4.39E+39 & T \\
31 & 111.7695 & 80.1694 & 0.210 & 0.008 & 5.09E-17 & 2.05E-18 & 1.53E+40 & 6.14E+38 & - & - & - & - & - & - & T \\
32 & 111.7826 & 80.1654 & 1.172 & 0.020 & 2.84E-16 & 4.84E-18 & 8.52E+40 & 1.45E+39 & - & - & - & - & - & - & L \\
33 & 111.8680 & 80.1499 & 0.723 & 0.016 & 1.75E-16 & 3.80E-18 & 5.26E+40 & 1.14E+39 & 0.089 & 0.005 & 4.69E-16 & 2.43E-17 & 9.02E+40 & 4.67E+39 & T \\
34 & 111.8289 & 80.1535 & 1.077 & 0.019 & 2.61E-16 & 4.64E-18 & 7.83E+40 & 1.39E+39 & 0.113 & 0.005 & 5.93E-16 & 2.73E-17 & 1.14E+41 & 5.26E+39 & T \\
35 & 111.8255 & 80.1521 & 0.801 & 0.017 & 1.94E-16 & 4.00E-18 & 5.82E+40 & 1.20E+39 & 0.106 & 0.005 & 5.57E-16 & 2.65E-17 & 1.07E+41 & 5.09E+39 & T \\
36 & 111.8157 & 80.1480 & 0.856 & 0.017 & 2.07E-16 & 4.14E-18 & 6.22E+40 & 1.24E+39 & 0.102 & 0.005 & 5.36E-16 & 2.60E-17 & 1.03E+41 & 4.99E+39 & T \\
37 & 111.7974 & 80.1534 & 0.451 & 0.012 & 1.09E-16 & 3.00E-18 & 3.28E+40 & 9.00E+38 & 0.022 & 0.002 & 1.15E-16 & 1.20E-17 & 2.21E+40 & 2.31E+39 & L \\
38 & 111.7703 & 80.1550 & 0.929 & 0.018 & 2.25E-16 & 4.31E-18 & 6.75E+40 & 1.29E+39 & 0.102 & 0.005 & 5.36E-16 & 2.60E-17 & 1.03E+41 & 4.99E+39 & T \\
39 & 111.7806 & 80.1554 & 0.991 & 0.018 & 2.40E-16 & 4.45E-18 & 7.20E+40 & 1.33E+39 & 0.109 & 0.005 & 5.72E-16 & 2.69E-17 & 1.10E+41 & 5.16E+39 & T \\
40 & 111.7766 & 80.1560 & 0.965 & 0.018 & 2.34E-16 & 4.39E-18 & 7.01E+40 & 1.32E+39 & 0.079 & 0.004 & 4.15E-16 & 2.29E-17 & 7.98E+40 & 4.40E+39 & L \\
41 & 111.7657 & 80.1584 & 0.931 & 0.018 & 2.26E-16 & 4.31E-18 & 6.77E+40 & 1.29E+39 & 0.097 & 0.005 & 5.07E-16 & 2.53E-17 & 9.74E+40 & 4.86E+39 & T \\
42 & 111.7524 & 80.1609 & 0.539 & 0.014 & 1.31E-16 & 3.28E-18 & 3.92E+40 & 9.84E+38 & 0.040 & 0.003 & 2.12E-16 & 1.64E-17 & 4.08E+40 & 3.14E+39 & T \\
43 & 111.7490 & 80.1627 & 0.585 & 0.014 & 1.42E-16 & 3.42E-18 & 4.25E+40 & 1.03E+39 & 0.063 & 0.004 & 3.33E-16 & 2.05E-17 & 6.40E+40 & 3.94E+39 & T \\
44 & 111.7446 & 80.1646 & 0.394 & 0.012 & 9.55E-17 & 2.81E-18 & 2.86E+40 & 8.41E+38 & - & - & - & - & - & - & L \\
45 & 111.7337 & 80.1683 & 0.623 & 0.015 & 1.51E-16 & 3.53E-18 & 4.53E+40 & 1.06E+39 & - & - & - & - & - & - & L \\
46 & 111.7836 & 80.1437 & 0.311 & 0.010 & 7.54E-17 & 2.49E-18 & 2.26E+40 & 7.47E+38 & 0.021 & 0.002 & 1.08E-16 & 1.16E-17 & 2.07E+40 & 2.24E+39 & T \\
47 & 111.7608 & 80.1488 & 0.533 & 0.013 & 1.29E-16 & 3.26E-18 & 3.87E+40 & 9.79E+38 & 0.093 & 0.005 & 4.91E-16 & 2.49E-17 & 9.44E+40 & 4.78E+39 & T \\
48 & 111.7812 & 80.1286 & 0.201 & 0.008 & 4.87E-17 & 2.00E-18 & 1.46E+40 & 6.01E+38 & - & - & - & - & - & - & L \\
49 & 111.7216 & 80.1377 & 0.256 & 0.009 & 6.21E-17 & 2.26E-18 & 1.86E+40 & 6.78E+38 & 0.027 & 0.003 & 1.41E-16 & 1.33E-17 & 2.70E+40 & 2.56E+39 & T \\
50 & 111.6904 & 80.1406 & 1.032 & 0.019 & 2.50E-16 & 4.54E-18 & 7.50E+40 & 1.36E+39 & 0.139 & 0.006 & 7.30E-16 & 3.03E-17 & 1.40E+41 & 5.83E+39 & T \\
\bottomrule
\end{tabular}
}
\end{table}
\end{landscape}
%%%%%%%%%%%%%%%%%%%%%%%%%%%%%%%%%%%%%%%%%%
\addtocounter{table}{-1}
\begin{landscape}
\begin{table}
\centering
\caption{continued}
% \label{Tab:sfknots}
\resizebox{1.3\textwidth}{!}{
\begin{tabular}{cccccccccccccccc}
\toprule
\multirow{2}{*}{ID} & \multirow{2}{*}{\thead{RA\\ {[}Degrees{]}}} & \multirow{2}{*}{\thead{DEC\\ {[}Degrees{]}}} & \multicolumn{6}{c}{NUV} & \multicolumn{6}{c}{FUV} & \multirow{2}{*}{Tight/Loose}\\ \cline{4-15}
 &  &  & \multicolumn{1}{c}{CPS} & \multicolumn{1}{c}{err} & \multicolumn{1}{c}{\thead{Flux\\ (erg/s/cm2/A)}} & \multicolumn{1}{c}{\thead{err\\ (erg/s/cm2/A)}} & \multicolumn{1}{c}{\thead{Log($L_{NUV}$)\\(erg/s)}} & \multicolumn{1}{c}{\thead{err\\(erg/s)}} & \multicolumn{1}{c}{CPS} & \multicolumn{1}{c}{err} & \multicolumn{1}{c}{\thead{Flux\\ (erg/s/cm2/A)}} & \multicolumn{1}{c}{\thead{err\\ (erg/s/cm2/A)}} & \multicolumn{1}{c}{\thead{Log($L_{FUV}$)\\(erg/s)}} & \multicolumn{1}{c}{\thead{err\\(erg/s)}} &  \\ 
 \midrule
51 & 111.7099 & 80.1468 & 1.019 & 0.019 & 2.47E-16 & 4.51E-18 & 7.41E+40 & 1.35E+39 & 0.144 & 0.006 & 7.56E-16 & 3.09E-17 & 1.45E+41 & 5.93E+39 & L \\
52 & 111.6657 & 80.1528 & 0.344 & 0.011 & 8.34E-17 & 2.62E-18 & 2.50E+40 & 7.86E+38 & 0.042 & 0.003 & 2.20E-16 & 1.67E-17 & 4.23E+40 & 3.20E+39 & T \\
53 & 111.6405 & 80.1535 & 0.167 & 0.008 & 4.05E-17 & 1.83E-18 & 1.21E+40 & 5.48E+38 & 0.021 & 0.002 & 1.12E-16 & 1.19E-17 & 2.16E+40 & 2.28E+39 & T \\
54 & 111.6170 & 80.1552 & 0.534 & 0.013 & 1.29E-16 & 3.27E-18 & 3.88E+40 & 9.79E+38 & - & - & - & - & - & - & T \\
55 & 111.6755 & 80.1573 & 0.318 & 0.010 & 7.71E-17 & 2.52E-18 & 2.31E+40 & 7.56E+38 & 0.043 & 0.003 & 2.25E-16 & 1.68E-17 & 4.33E+40 & 3.24E+39 & T \\
56 & 111.7138 & 80.1567 & 0.445 & 0.012 & 1.08E-16 & 2.98E-18 & 3.23E+40 & 8.94E+38 & 0.040 & 0.003 & 2.12E-16 & 1.63E-17 & 4.07E+40 & 3.14E+39 & T \\
57 & 111.7260 & 80.1571 & 0.301 & 0.010 & 7.30E-17 & 2.45E-18 & 2.19E+40 & 7.35E+38 & 0.031 & 0.003 & 1.63E-16 & 1.43E-17 & 3.14E+40 & 2.76E+39 & L \\
58 & 111.7125 & 80.1640 & 0.290 & 0.010 & 7.03E-17 & 2.41E-18 & 2.11E+40 & 7.22E+38 & 0.033 & 0.003 & 1.73E-16 & 1.48E-17 & 3.33E+40 & 2.84E+39 & L \\
59 & 111.7006 & 80.1638 & 0.092 & 0.006 & 2.24E-17 & 1.36E-18 & 6.72E+39 & 4.07E+38 & 0.010 & 0.002 & 5.08E-17 & 8.00E-18 & 9.76E+39 & 1.54E+39 & L \\
60 & 111.7061 & 80.1660 & 0.147 & 0.007 & 3.56E-17 & 1.71E-18 & 1.07E+40 & 5.14E+38 & 0.011 & 0.002 & 5.99E-17 & 8.69E-18 & 1.15E+40 & 1.67E+39 & T \\
61 & 111.6684 & 80.1680 & 0.576 & 0.014 & 1.40E-16 & 3.39E-18 & 4.19E+40 & 1.02E+39 & 0.074 & 0.004 & 3.90E-16 & 2.22E-17 & 7.50E+40 & 4.26E+39 & T \\
62 & 111.6649 & 80.1728 & 0.460 & 0.013 & 1.12E-16 & 3.03E-18 & 3.34E+40 & 9.09E+38 & 0.058 & 0.004 & 3.05E-16 & 1.96E-17 & 5.87E+40 & 3.77E+39 & T \\
63 & 111.6885 & 80.1711 & 2.924 & 0.032 & 7.09E-16 & 7.64E-18 & 2.13E+41 & 2.29E+39 & 0.389 & 0.010 & 2.04E-15 & 5.07E-17 & 3.93E+41 & 9.75E+39 & T \\
64 & 111.6965 & 80.1793 & 1.309 & 0.021 & 3.17E-16 & 5.11E-18 & 9.51E+40 & 1.53E+39 & 0.204 & 0.007 & 1.07E-15 & 3.67E-17 & 2.06E+41 & 7.06E+39 & T \\
65 & 111.7056 & 80.1833 & 0.276 & 0.010 & 6.69E-17 & 2.35E-18 & 2.01E+40 & 7.04E+38 & 0.049 & 0.003 & 2.55E-16 & 1.79E-17 & 4.90E+40 & 3.44E+39 & L \\
66 & 111.7152 & 80.1878 & 0.430 & 0.012 & 1.04E-16 & 2.93E-18 & 3.13E+40 & 8.79E+38 & 0.049 & 0.003 & 2.59E-16 & 1.81E-17 & 4.98E+40 & 3.47E+39 & L \\
67 & 111.6731 & 80.1881 & 0.560 & 0.014 & 1.36E-16 & 3.34E-18 & 4.07E+40 & 1.00E+39 & 0.056 & 0.004 & 2.93E-16 & 1.92E-17 & 5.63E+40 & 3.69E+39 & L \\
68 & 111.6299 & 80.1834 & 1.607 & 0.023 & 3.90E-16 & 5.67E-18 & 1.17E+41 & 1.70E+39 & 0.313 & 0.009 & 1.64E-15 & 4.55E-17 & 3.16E+41 & 8.75E+39 & T \\
69 & 111.6406 & 80.1886 & 1.431 & 0.022 & 3.47E-16 & 5.35E-18 & 1.04E+41 & 1.60E+39 & 0.171 & 0.006 & 8.98E-16 & 3.36E-17 & 1.73E+41 & 6.47E+39 & L \\
70 & 111.6624 & 80.1922 & 0.307 & 0.010 & 7.44E-17 & 2.48E-18 & 2.23E+40 & 7.43E+38 & 0.040 & 0.003 & 2.11E-16 & 1.63E-17 & 4.06E+40 & 3.13E+39 & L \\
71 & 111.6905 & 80.1956 & 0.599 & 0.014 & 1.45E-16 & 3.46E-18 & 4.35E+40 & 1.04E+39 & 0.075 & 0.004 & 3.94E-16 & 2.23E-17 & 7.57E+40 & 4.28E+39 & L \\
72 & 111.7011 & 80.2083 & 0.466 & 0.013 & 1.13E-16 & 3.05E-18 & 3.39E+40 & 9.15E+38 & 0.064 & 0.004 & 3.37E-16 & 2.06E-17 & 6.48E+40 & 3.96E+39 & T \\
73 & 111.7522 & 80.2002 & 0.493 & 0.013 & 1.19E-16 & 3.14E-18 & 3.58E+40 & 9.41E+38 & 0.049 & 0.003 & 2.59E-16 & 1.81E-17 & 4.99E+40 & 3.47E+39 & L \\
74 & 111.7773 & 80.2001 & 1.742 & 0.024 & 4.22E-16 & 5.90E-18 & 1.27E+41 & 1.77E+39 & 0.213 & 0.007 & 1.12E-15 & 3.75E-17 & 2.15E+41 & 7.22E+39 & L \\
75 & 111.7619 & 80.2027 & 0.664 & 0.015 & 1.61E-16 & 3.64E-18 & 4.83E+40 & 1.09E+39 & 0.069 & 0.004 & 3.60E-16 & 2.13E-17 & 6.92E+40 & 4.09E+39 & L \\
76 & 111.7478 & 80.2046 & 0.497 & 0.013 & 1.20E-16 & 3.15E-18 & 3.61E+40 & 9.45E+38 & 0.075 & 0.004 & 3.96E-16 & 2.23E-17 & 7.61E+40 & 4.29E+39 & L \\
77 & 111.7493 & 80.2073 & 0.126 & 0.007 & 3.05E-17 & 1.59E-18 & 9.16E+39 & 4.76E+38 & - & - & - & - & - & - & T \\
78 & 111.7846 & 80.1916 & 0.494 & 0.013 & 1.20E-16 & 3.14E-18 & 3.59E+40 & 9.42E+38 & 0.047 & 0.003 & 2.49E-16 & 1.77E-17 & 4.78E+40 & 3.40E+39 & T \\
\bottomrule
 \end{tabular}
}
\end{table}
\end{landscape}
%%%%%%%%%%%%%%%%%%%%%%%%%%%%%%%%%%%%%%%%%%%
\begin{landscape}
\begin{table}
\centering
\caption{Derived properties of the star forming knots}
\label{Tab:sfknots2}
\resizebox{1.3\textwidth}{!}{
\begin{tabular}{cccccccccccccccc}
\toprule
\multirow{2}{*}{ID} & \multirow{2}{*}{\thead{r\\(arcsec)}} & \multirow{2}{*}{\thead{r\\(kpc)}} & \multirow{2}{*}{\thead{r$^{\prime}$\\(arcsec)}} & \multirow{2}{*}{\thead{r$^{\prime}$\\(kpc)}} & \multicolumn{4}{c}{NUV} & \multicolumn{4}{c}{FUV} & \multirow{2}{*}{\thead{Age\\(Myrs)}} & \multirow{2}{*}{\thead{Mass\\($M_{\odot}$)}} & \multirow{2}{*}{\thead{metallicity\\Z=0.02(solar)}} \\ \cline{6-13}
 &  &  &  &  & \thead{FWHM\\(arcsec)} & \thead{FWHM\\(pc)} & \thead{SFR\\($10^{-3}M_{\odot}/yr$)} & \thead{SFR\_err\\($10^{-3}M_{\odot}/yr$)}  & \thead{FWHM\\(arcsec)} & \thead{FWHM\\(pc)} & \thead{SFR\\($10^{-3}M_{\odot}/yr$)} & \thead{SFR\_err\\($10^{-3}M_{\odot}/yr$)} &  &  &  \\ 
 \midrule
1 & 151.48 & 23.63 & 162.29 & 25.32 & 2.14 & 333.84 & 5.59 & 0.12 & 2.45 & 382.2 & 5.21 & 0.24 & 42.089 & 1.01E+06 & 0.0004 \\
2 & 149.87 & 23.38 & 151.50 & 23.63 & 2.91 & 453.96 & 4.09 & 0.10 & 2.88 & 449.28 & 3.67 & 0.20 & 56.517 & 9.74E+05 & 0.004 \\
3 & 152.00 & 23.71 & 152.39 & 23.77 & 4.34 & 677.04 & 7.43 & 0.14 & - & - & - & - & - & - & - \\
4 & 140.90 & 21.98 & 141.32 & 22.05 & 2.9 & 452.4 & 9.91 & 0.16 & 2.46 & 383.76 & 7.74 & 0.30 & 26.07 & 1.04E+06 & 0.008 \\
5 & 132.58 & 20.68 & 132.87 & 20.73 & 2.42 & 377.52 & 3.93 & 0.10 & - & - & - & - & - & - & - \\
6 & 131.11 & 20.45 & 132.20 & 20.62 & 3.16 & 492.96 & 6.59 & 0.13 & - & - & - & - & - & - & - \\
7 & 114.31 & 17.83 & 117.30 & 18.30 & 3.07 & 478.92 & 5.79 & 0.12 & 3.12 & 486.72 & 4.79 & 0.23 & 46.15 & 1.02E+06 & 0.0004 \\
8 & 110.03 & 17.16 & 125.86 & 19.63 & 2.59 & 404.04 & 2.78 & 0.08 & - & - & - & - & - & - & - \\
9 & 99.73 & 15.56 & 115.20 & 17.97 & 2.5 & 390 & 2.89 & 0.09 & - & - & - & - & - & - & - \\
10 & 103.93 & 16.21 & 124.20 & 19.38 & 2.05 & 319.8 & 0.75 & 0.04 & 1.63 & 254.28 & 0.54 & 0.08 & 100.04 & 6.55E+05 & 0.05 \\
11 & 88.76 & 13.85 & 122.56 & 19.12 & 3.9 & 608.4 & 3.28 & 0.09 & 4.43 & 691.08 & 3.11 & 0.19 & 64.30 & 9.63E+05 & 0.004 \\
12 & 66.99 & 10.45 & 97.80 & 15.26 & 2.21 & 344.76 & 7.23 & 0.13 & 3.85 & 600.6 & 9.20 & 0.32 & 26.07 & 1.00E+06 & 0.0004 \\
13 & 55.75 & 8.70 & 69.99 & 10.92 & 2.28 & 355.68 & 5.69 & 0.12 & 1.9 & 296.4 & 3.92 & 0.21 & 36.32 & 9.97E+05 & 0.02 \\
14 & 72.17 & 11.26 & 114.40 & 17.85 & 1.74 & 271.44 & 0.81 & 0.05 & - & - & - & - & - & - & - \\
15 & 87.07 & 13.58 & 134.81 & 21.03 & 2.46 & 383.76 & 3.24 & 0.09 & 2.53 & 394.68 & 2.52 & 0.17 & 78.02 & 9.89E+05 & 0.0004 \\
16 & 106.86 & 16.67 & 160.14 & 24.98 & 1.63 & 254.28 & 2.23 & 0.07 & - & - & - & - & - & - & - \\
17 & 121.32 & 18.93 & 179.55 & 28.01 & 1.25 & 195 & 1.09 & 0.05 & - & - & - & - & - & - & - \\
18 & 68.18 & 10.64 & 102.46 & 15.98 & 2.34 & 365.04 & 1.52 & 0.06 & - & - & - & - & - & - & - \\
19 & 77.78 & 12.13 & 105.19 & 16.41 & 2.65 & 413.4 & 1.60 & 0.06 & - & - & - & - & - & - & - \\
20 & 62.02 & 9.68 & 78.08 & 12.18 & 2.18 & 340.08 & 8.17 & 0.14 & 2.73 & 425.88 & 6.90 & 0.28 & 20.14 & 9.62E+05 & 0.008 \\
21 & 53.82 & 8.40 & 61.55 & 9.60 & 3.378 & 526.968 & 12.78 & 0.18 & 4.17 & 650.52 & 10.14 & 0.34 & 32.22 & 9.71E+05 & 0.004 \\
22 & 81.76 & 12.75 & 98.16 & 15.31 & 2.68 & 418.08 & 2.31 & 0.08 & 2.73 & 425.88 & 1.68 & 0.14 & 82.45 & 1.00E+06 & 0.008 \\
23 & 50.04 & 7.81 & 66.85 & 10.43 & 1.85 & 288.6 & 1.62 & 0.06 & 2.26 & 352.56 & 1.31 & 0.12 & 96.42 & 9.73E+05 & 0.008 \\
24 & 37.46 & 5.84 & 51.66 & 8.06 & 1.82 & 283.92 & 1.42 & 0.06 & - & - & - & - & - & - & - \\
25 & 35.77 & 5.58 & 56.10 & 8.75 & 2.04 & 318.24 & 0.82 & 0.05 & - & - & - & - & - & - & - \\
26 & 28.35 & 4.42 & 44.90 & 7.00 & 1.56 & 243.36 & 9.78 & 0.16 & - & - & - & - & - & - & - \\
27 & 32.81 & 5.12 & 40.90 & 6.38 & 2.49 & 388.44 & 2.55 & 0.08 & - & - & - & - & - & - & - \\
28 & 24.28 & 3.79 & 34.38 & 5.36 & 2.12 & 330.72 & 4.05 & 0.10 & 2.25 & 351 & 2.04 & 0.15 & 56.52 & 1.01E+06 & 0.05 \\
29 & 26.58 & 4.15 & 31.01 & 4.84 & 1.54 & 240.24 & 2.10 & 0.07 & 2.37 & 369.72 & 1.33 & 0.12 & 98.22 & 1.01E+06 & 0.008 \\
30 & 29.95 & 4.67 & 30.92 & 4.82 & 2.42 & 377.52 & 5.07 & 0.11 & 3.16 & 492.96 & 3.69 & 0.20 & 46.15 & 1.02E+06 & 0.008 \\
31 & 32.40 & 5.05 & 32.43 & 5.06 & 1.86 & 290.16 & 1.55 & 0.06 & - & - & - & - & - & - & - \\
32 & 47.79 & 7.46 & 48.47 & 7.56 & 3.56 & 555.36 & 8.65 & 0.15 & - & - & - & - & - & - & - \\
33 & 120.29 & 18.77 & 131.56 & 20.52 & 2.45 & 382.2 & 5.34 & 0.12 & 2.73 & 425.88 & 4.20 & 0.22 & 52.02 & 1.01E+06 & 0.0004 \\
34 & 97.71 & 15.24 & 102.69 & 16.02 & 3.64 & 567.84 & 7.95 & 0.14 & 4.31 & 672.36 & 5.30 & 0.24 & 20.14 & 9.44E+05 & 0.05 \\
35 & 101.26 & 15.80 & 105.50 & 16.46 & 2.59 & 404.04 & 5.91 & 0.12 & 2.48 & 386.88 & 4.98 & 0.24 & 44.07 & 1.01E+06 & 0.004 \\
36 & 113.40 & 17.69 & 115.98 & 18.09 & 2.78 & 433.68 & 6.32 & 0.13 & 2.86 & 446.16 & 4.79 & 0.23 & 36.32 & 9.92E+05 & 0.008 \\
37 & 91.86 & 14.33 & 93.12 & 14.53 & 2.24 & 349.44 & 3.33 & 0.09 & 2.16 & 336.96 & 1.03 & 0.11 & 92.08 & 1.10E+06 & 0.05 \\
38 & 84.06 & 13.11 & 84.09 & 13.12 & 2.87 & 447.72 & 6.86 & 0.13 & 3.92 & 611.52 & 4.79 & 0.23 & 30.21 & 9.69E+05 & 0.02 \\
39 & 83.13 & 12.97 & 83.42 & 13.01 & 2.15 & 335.4 & 7.32 & 0.14 & 2.5 & 390 & 5.12 & 0.24 & 30.21 & 1.04E+06 & 0.02 \\
40 & 80.71 & 12.59 & 80.86 & 12.61 & 4.91 & 765.96 & 7.12 & 0.13 & 3.8 & 592.8 & 3.71 & 0.20 & 28.07 & 9.90E+05 & 0.05 \\
41 & 71.99 & 11.23 & 71.99 & 11.23 & 3.02 & 471.12 & 6.87 & 0.13 & 3.71 & 578.76 & 4.53 & 0.23 & 24.22 & 1.04E+06 & 0.05 \\
42 & 63.43 & 9.89 & 63.77 & 9.95 & 2.29 & 357.24 & 3.98 & 0.10 & 1.77 & 276.12 & 1.90 & 0.15 & 60.28 & 1.04E+06 & 0.05 \\
43 & 57.27 & 8.93 & 57.85 & 9.02 & 2.77 & 432.12 & 4.32 & 0.10 & 2.65 & 413.4 & 2.97 & 0.18 & 46.15 & 1.01E+06 & 0.02 \\
44 & 51.42 & 8.02 & 52.46 & 8.18 & 3.32 & 517.92 & 2.91 & 0.09 & - & - & - & - & - & - & - \\
45 & 41.33 & 6.45 & 44.55 & 6.95 & 3.42 & 533.52 & 4.60 & 0.11 & - & - & - & - & - & - & - \\
46 & 125.22 & 19.53 & 125.50 & 19.58 & 2.3 & 358.8 & 2.30 & 0.08 & 2.58 & 402.48 & 0.96 & 0.10 & 92.08 & 1.03E+06 & 0.05 \\
47 & 106.59 & 16.63 & 106.61 & 16.63 & 3.02 & 471.12 & 3.93 & 0.10 & 4.07 & 634.92 & 4.39 & 0.22 & 46.15 & 9.32E+05 & 0.0004 \\
48 & 179.54 & 28.01 & 179.68 & 28.03 & 1.96 & 305.76 & 1.48 & 0.06 & - & - & - & - & - & - & - \\
49 & 149.09 & 23.26 & 150.62 & 23.50 & 1.93 & 301.08 & 1.89 & 0.07 & 2.38 & 371.28 & 1.26 & 0.12 & 100.04 & 9.83E+05 & 0.008 \\
50 & 143.88 & 22.45 & 148.64 & 23.19 & 3.05 & 475.8 & 7.62 & 0.14 & 4.55 & 709.8 & 6.52 & 0.27 & 34.05 & 9.78E+05 & 0.004 \\
\bottomrule
\end{tabular}
}
\vspace{-12.68889pt}
\end{table}
\end{landscape}
\addtocounter{table}{-1}
%%%%%%%%%%%%%%%%%%%%%%%%%%%%%%%%%%%%%%%%%
\begin{landscape}
\begin{table}
\centering
\caption{continued}
% \label{Tab:sfknots2}
\resizebox{1.3\textwidth}{!}{
\begin{tabular}{cccccccccccccccc}
\toprule
\multirow{2}{*}{ID} & \multirow{2}{*}{\thead{r\\(arcsec)}} & \multirow{2}{*}{\thead{r\\(kpc)}} & \multirow{2}{*}{\thead{r$^{\prime}$\\(arcsec)}} & \multirow{2}{*}{\thead{r$^{\prime}$\\(kpc)}} & \multicolumn{4}{c}{NUV} & \multicolumn{4}{c}{FUV} & \multirow{2}{*}{\thead{Age\\(Myrs)}} & \multirow{2}{*}{\thead{Mass\\($M_{\odot}$)}} & \multirow{2}{*}{\thead{metallicity\\Z=0.02(solar)}} \\ \cline{6-13}
 &  &  &  &  & \thead{FWHM\\(arcsec)} & \thead{FWHM\\(pc)} & \thead{SFR\\($10^{-3}M_{\odot}/yr$)} & \thead{SFR\_err\\($10^{-3}M_{\odot}/yr$)}  & \thead{FWHM\\(arcsec)} & \thead{FWHM\\(pc)} & \thead{SFR\\($10^{-3}M_{\odot}/yr$)} & \thead{SFR\_err\\($10^{-3}M_{\odot}/yr$)} &  &  &  \\ 
 \midrule
51 & 118.80 & 18.53 & 121.94 & 19.02 & 3.22 & 502.32 & 7.52 & 0.14 & 3.79 & 591.24 & 6.76 & 0.28 & 34.05 & 1.01E+06 & 0.0004 \\
52 & 110.72 & 17.27 & 122.81 & 19.16 & 3.02 & 471.12 & 2.54 & 0.08 & 3.47 & 541.32 & 1.97 & 0.15 & 96.42 & 9.97E+05 & 0.004 \\
53 & 118.31 & 18.46 & 137.22 & 21.41 & 2.21 & 344.76 & 1.23 & 0.06 & 3.71 & 578.76 & 1.00 & 0.11 & 98.22 & 9.67E+05 & 0.02 \\
54 & 123.96 & 19.34 & 151.40 & 23.62 & 1.67 & 260.52 & 3.94 & 0.10 & - & - & - & - & - & - & - \\
55 & 94.18 & 14.69 & 106.00 & 16.54 & 1.92 & 299.52 & 2.35 & 0.08 & 2.71 & 422.76 & 2.01 & 0.15 & 94.66 & 9.97E+05 & 0.0004 \\
56 & 84.34 & 13.16 & 88.27 & 13.77 & 2.56 & 399.36 & 3.28 & 0.09 & 3.5 & 546 & 1.89 & 0.15 & 58.10 & 9.81E+05 & 0.05 \\
57 & 80.29 & 12.52 & 82.66 & 12.89 & 2.52 & 393.12 & 2.22 & 0.07 & 2.63 & 410.28 & 1.46 & 0.13 & 90.40 & 9.90E+05 & 0.008 \\
58 & 61.41 & 9.58 & 67.53 & 10.54 & 2.9 & 452.4 & 2.14 & 0.07 & 2.72 & 424.32 & 1.55 & 0.13 & 86.33 & 9.85E+05 & 0.008 \\
59 & 66.07 & 10.31 & 74.96 & 11.69 & 1.68 & 262.08 & 0.68 & 0.04 & 2.16 & 336.96 & 0.45 & 0.07 & 100.04 & 5.57E+05 & 0.05 \\
60 & 57.67 & 9.00 & 66.41 & 10.36 & 1.66 & 258.96 & 1.09 & 0.05 & 2.42 & 377.52 & 0.54 & 0.08 & 100.04 & 6.57E+05 & 0.05 \\
61 & 70.61 & 11.02 & 93.92 & 14.65 & 2.05 & 319.8 & 4.25 & 0.10 & 2.53 & 394.68 & 3.49 & 0.20 & 60.28 & 9.97E+05 & 0.0004 \\
62 & 65.28 & 10.18 & 96.87 & 15.11 & 2.48 & 386.88 & 3.40 & 0.09 & 2.84 & 443.04 & 2.73 & 0.18 & 72.47 & 9.82E+05 & 0.004 \\
63 & 54.32 & 8.47 & 74.16 & 11.57 & 2.69 & 419.64 & 21.58 & 0.23 & 3.18 & 496.08 & 18.26 & 0.45 & 14.07 & 1.06E+06 & 0.008 \\
64 & 42.80 & 6.68 & 67.62 & 10.55 & 2.21 & 344.76 & 9.66 & 0.16 & 2.87 & 447.72 & 9.58 & 0.33 & 24.22 & 9.48E+05 & 0.004 \\
65 & 41.02 & 6.40 & 57.56 & 8.98 & 3.44 & 536.64 & 2.04 & 0.07 & 3.23 & 503.88 & 2.28 & 0.16 & 80.20 & 9.26E+05 & 0.0004 \\
66 & 45.95 & 7.17 & 54.07 & 8.43 & 3.01 & 469.56 & 3.17 & 0.09 & 2.98 & 464.88 & 2.32 & 0.16 & 84.76 & 1.01E+06 & 0.0004 \\
67 & 66.96 & 10.45 & 89.26 & 13.93 & 3.19 & 497.64 & 4.13 & 0.10 & 3.27 & 510.12 & 2.62 & 0.17 & 44.07 & 1.00E+06 & 0.05 \\
68 & 85.49 & 13.34 & 131.27 & 20.48 & 2.32 & 361.92 & 11.86 & 0.17 & 3.95 & 616.2 & 14.69 & 0.41 & 18.04 & 1.01E+06 & 0.004 \\
69 & 85.38 & 13.32 & 119.61 & 18.66 & 4.48 & 698.88 & 10.56 & 0.16 & 5.19 & 809.64 & 8.03 & 0.30 & 24.22 & 9.78E+05 & 0.008 \\
70 & 80.82 & 12.61 & 102.06 & 15.92 & 2.98 & 464.88 & 2.27 & 0.08 & 3.07 & 478.92 & 1.89 & 0.15 & 98.22 & 9.81E+05 & 0.004 \\
71 & 77.36 & 12.07 & 87.39 & 13.63 & 2.61 & 407.16 & 4.42 & 0.11 & 3.35 & 522.6 & 3.52 & 0.20 & 60.28 & 1.01E+06 & 0.0004 \\
72 & 114.76 & 17.90 & 119.16 & 18.59 & 2.2 & 343.2 & 3.44 & 0.09 & 2.93 & 457.08 & 3.01 & 0.18 & 66.10 & 9.66E+05 & 0.004 \\
73 & 79.17 & 12.35 & 79.45 & 12.39 & 4.13 & 644.28 & 3.64 & 0.10 & 4.19 & 653.64 & 2.32 & 0.16 & 50.14 & 9.90E+05 & 0.05 \\
74 & 78.56 & 12.25 & 78.75 & 12.29 & 4.17 & 650.52 & 12.86 & 0.18 & 4.85 & 756.6 & 10.00 & 0.34 & 20.14 & 9.49E+05 & 0.008 \\
75 & 87.57 & 13.66 & 87.59 & 13.66 & 3.51 & 547.56 & 4.90 & 0.11 & 5.16 & 804.96 & 3.22 & 0.19 & 34.05 & 1.02E+06 & 0.05 \\
76 & 95.03 & 14.82 & 95.43 & 14.89 & 2.58 & 402.48 & 3.67 & 0.10 & 2.78 & 433.68 & 3.54 & 0.20 & 58.10 & 9.71E+05 & 0.0004 \\
77 & 104.74 & 16.34 & 105.05 & 16.39 & 1.58 & 246.48 & 0.93 & 0.05 & - & - & - & - & - & - & - \\
78 & 48.98 & 7.64 & 49.80 & 7.77 & 2.21 & 344.76 & 3.65 & 0.10 & 2.36 & 368.16 & 2.22 & 0.16 & 52.02 & 9.96E+05 & 0.05 \\
\bottomrule
 \end{tabular}
}
\end{table}
\end{landscape}
%%%%%%%%%%%%%%%%%%%%%%%%%%%%%%%%%%%%%%%%%end
%%%%%%%%%%%%%%%%%%%%%%%%%%%%%%%%%%%%%%%%%%%%%%%%%%
\bibliographystyle{mnras} 
\bibliography{ref}
\label{lastpage}
\end{document}